\documentclass[usenatbib]{aa}
\usepackage{graphicx}
\usepackage{graphics}
\usepackage{txfonts}
\usepackage{epsfig}
\usepackage{natbib}
\usepackage[]{color}

\newcommand{\placetabone}{
\renewcommand{\tabcolsep}{4pt}
\begin{table*}
\caption{Main properties of the F3D galaxies.}
\centering
\begin{tabular}{lcccccccccc}
\hline\hline
\noalign{\smallskip}
Object & R.A.   & Dec.  & Type       & $cz$ & $m_B$ & $R_{\rm e}$  & $R_{25}$ & $\mu_B(30\;{\rm arcsec})$  & Alternative Names \\
       & [$^{\rm h}\;^{\rm m}\;^{\rm s}$]      & [$^\circ\;'\;''$]     &            & [km~s$^{-1}$] & [mag]   & [arcmin] & [arcmin]   & [mag~arcsec$^{-2}$] &                   \\
(1)    & (2)        & (3)       & (4)        & (5)  & (6)   & (7)    & (8)      & (9)            & (10)              \\
\noalign{\smallskip}
\hline
\noalign{\smallskip}
FCC~083 & 03 30 35.1 & $-34$ 51 14 & E5         & 1543 & 12.3 & 0.46 & 1.5$^{a}$ & 22.7 & NGC~1351, ESO~358-G21 \\
FCC~090 & 03 31 08.1 & $-36$ 17 19 & E4 pec     & 1916 & 15.0 & 0.20 & 0.5$^{a}$ & 25.1 & ... \\
FCC~113 & 03 33 06.8 & $-34$ 48 26 & ScdIII pec & 1553 & 15.2 & 0.25 & 0.6$^{b}$ & ...  & ESO~358-015 \\
FCC~119 & 03 33 33.7 & $-33$ 34 18 & S0 pec     & 1554 & 15.0 & 0.29 & 0.4$^{a}$ & 25.5 & ... \\
FCC~143 & 03 34 59.1 & $-35$ 10 10 & E3         & 1376 & 14.3 & 0.16 & 0.5$^{a}$ & 25.0 & NGC~1373, ESO~358-G21 \\
FCC~147 & 03 35 16.8 & $-35$ 13 34 & E0         & 1386 & 11.9 & 0.42 & 1.3$^{a}$ & 22.6 & NGC~1374, ESO~358-G23 \\
FCC~148 & 03 35 16.8 & $-35$ 15 56 & S0         &  730 & 13.6 & 0.25 & 1.1$^{a}$ & 22.7 & NGC~1375, ESO~358-G24 \\
FCC~153 & 03 35 30.9 & $-34$ 26 45 & S0         & 1639 & 13.0 & 0.19 & 1.3$^{b}$ & 21.2 & IC~1963, ESO~358-G26 \\
FCC~161 & 03 36 04.0 & $-35$ 26 30 & E0         & 1405 & 11.7 & 0.39 & 2.7$^{a}$ & 22.6 & NGC~1379, ESO~358-G37 \\
FCC~167 & 03 36 27.5 & $-34$ 58 31 & S0/a       & 1827 & 11.3 & 0.62 & 2.0$^{a}$ & 21.1 & NGC~1380, ESO~358-G28 \\
FCC~170 & 03 36 31.6 & $-35$ 17 43 & S0         & 1793 & 13.0 & 0.23 & 1.3$^{a}$ & 21.2 & NGC~1381, ESO~358-G29 \\
FCC~176 & 03 36 45.0 & $-36$ 15 17 & SBa        & 1465 & 13.7 & 0.43 & 0.8$^{b}$ & ...  & NGC~1369, ESO~358-034 \\
FCC~177 & 03 36 47.4 & $-34$ 44 17 & S0         & 1495 & 13.2 & 0.25 & 1.2$^{a}$ & 21.6 & NGC~1380A, ESO~358-G33 \\
FCC~179 & 03 36 46.3 & $-35$ 59 57 & Sa         &  839 & 12.4 & ...  & 1.7$^{b}$ &...   & NGC~1386, ESO~358-035 \\
FCC~182 & 03 36 54.3 & $-35$ 22 23 & SB0 pec    & 1823 & 14.9 & 0.19 & 0.6$^{b}$ & $\geq 25.0$ & ... \\
FCC~184 & 03 36 56.9 & $-35$ 30 24 & SB0        & 1337 & 12.3 & 0.83 & 1.8$^{b}$ & 21.1 & NGC~1387, ESO~358-G36 \\
FCC~190 & 03 37 08.9 & $-35$ 11 37 & SB0        & 1784 & 13.5 & 0.27 & 0.8$^{b}$ & 24.0 & NGC~1380B, ESO~358-G37 \\
FCC~193 & 03 37 11.7 & $-35$ 44 40 & SB0        &  999 & 12.8 & 0.33 & 1.4$^{b}$ & 21.9 & NGC~1389, ESO~358-G38 \\
FCC~213 & 03 38 29.2 & $-35$ 27 02 & E0         & 1430 & 10.6 & 2.12 & 3.5$^{a}$ & 21.4 & NGC~1399, ESO~358-G45 \\
FCC~219 & 03 38 52.1 & $-35$ 35 38 & E2         & 1944 & 10.9 & 0.41 & 2.1$^{a}$ & 21.6 & NGC~1404, ESO~358-G46 \\
FCC~249 & 03 40 41.9 & $-37$ 30 33 & E0         & 1521 & 13.6 & 0.16 & 0.6$^{a}$ & 24.7 & NGC~1419, ESO~301-G23 \\
FCC~255 & 03 41 03.4 & $-33$ 46 38 & S0         & 1217 & 13.7 & 0.23 & 1.0$^{b}$ & 20.4 & ESO~358-G50 \\
FCC~263 & 03 41 32.2 & $-34$ 53 17 & SBcdIII    & 1708 & 14.6 & 0.34 & 0.8$^{b}$ & ...  & ESO~358-051 \\
FCC~267 & 03 41 45.4 & $-33$ 47 24 & SmIV       & -    & 16.0 & 0.20 & 1.1$^{b}$ & ...  & ... \\
FCC~276 & 03 42 19.2 & $-35$ 23 36 & E4         & 1454 & 11.8 & 0.54 & 1.6$^{a}$ & 22.2 & NGC~1427, ESO~358-G52 \\
FCC~277 & 03 42 22.6 & $-35$ 09 10 & E5         & 1620 & 13.8 & 0.16 & 0.7$^{a}$ & 24.1 & NGC~1428, ESO~358-G53 \\
FCC~285 & 03 43 01.8 & $-36$ 16 11 & SdIII      & 895  & 14.2 & 0.48 & 1.0$^{b}$ & ...  & NGC~1437A, ESO~358-054 \\
FCC~290 & 03 43 37.0 & $-35$ 51 13 & ScII       & 1217 & 12.8 & ...  & 1.5$^{b}$ & ...  & NGC~1436, ESO~358-058 \\
FCC~301 & 03 45 03.5 & $-35$ 58 17 & E4         & 1020 & 14.2 & 0.17 & 0.5$^{a}$ & 24.9 & ESO~358-G59 \\
FCC~306 & 03 45 45.3 & $-36$ 20 40 & SBmIII     & -    & 15.6 & ...  & 0.3$^{b}$ & ...  & ... \\
FCC~308 & 03 45 54.7 & $-36$ 21 25 & Sd         & -    & 13.8 & 0.06 & 1.3$^{b}$ & ...  & NGC~1437B, ESO~358-061 \\
FCC~310 & 03 46 13.7 & $-36$ 41 43 & SB0        & 1352 & 13.5 & 0.43 & 0.9$^{b}$ & 20.0 & NGC~1460, ESO~358-G62 \\
FCC~312 & 03 46 18.9 & $-34$ 56 31 & Scd        & 1932 & 13.5 & ...  & 2.4$^{b}$ & ...  & ESO~358-063 \\
\noalign{\smallskip}
\hline
\end{tabular}
\tablefoot{(1) Galaxy name from \citet{Ferguson1989}. (2),
  (3) Right ascension and declination (J2000.0). (4), (5), (6), and
  (7) Morphological type, heliocentric radial velocity, total $B$-band
  magnitude, and $B$-band effective radius from
  \citet{Ferguson1989}. (8) Semi-major axis of the best-fitting
  ellipse to the isophote at $\mu_B = 25$ mag~arcsec$^{-2}$ from
  $^{(a)}$\citet{Caon1994} or $^{(b)}$\citet{RC3}. (9) $B$-band
  surface brightness level at $R=30$ arcsec from
  \citet{Caon1994}. (10) Galaxy alternative names.}
\label{tab:sample}
\end{table*}
}

\newcommand{\placetabtwo}{
\renewcommand{\tabcolsep}{4pt}
\begin{table*}
\caption{\label{tab:log} MUSE pointings of the F3D galaxies.}
\centering
\begin{tabular}{lcccc}
\hline\hline
\noalign{\smallskip}
Object & Central pointing & Middle pointing & Halo pointing & Date\\
           &      [s]        & [s]     &       [s] &   [mmm yyyy]   \\
(1) & (2) & (3)  & (4) & (5)\\
\noalign{\smallskip}
\hline
\noalign{\smallskip}
FCC~083 & $5\times720$ & ...          & $9\times600$ & Oct 2016; Dec 2016; Nov 2017\\
FCC~090 & $5\times720$ & ...          & $3\times600$ & Dec 2016; Nov 2017\\
FCC~113 & $4\times900$ & ...          & ...          & Jul 2017; Nov 2017\\
FCC~119 & $5\times720$ & ...          & $3\times600$ & Nov 2017\\
FCC~143 & $5\times720$ & ...          & $3\times600$ & Nov--Dec 2017\\
FCC~147 & $5\times720$ & ...          & $9\times600$ & Nov--Dec 2017\\
FCC~148 & $5\times720$ & ...          & $9\times600$ & Dec 2016\\
FCC~153 & $5\times720$ & ...          & $9\times600$ & Nov--Dec 2017\\
FCC~161 & $5\times720$ & ...          & $9\times600$ & Dec 2017\\
FCC~167 & $5\times720$ & $6\times600$ & $9\times600$ & Dec 2016; Jan 2017; Nov 2017\\
FCC~170 & $5\times720$ & ...          & $12\times600$ & Dec 2016; Jan 2017\\
FCC~176 & $4\times900$ & ...          & ...          & Sep 2017; Dec 2017\\
FCC~177 & $5\times720$ & ...          & $9\times600$ & Nov--Dec 2017\\
FCC~179 & $4\times900$ & ...          & $4\times900$ & Oct--Dec 2017\\
FCC~182 & $5\times720$ & ...          & $3\times600$ & Jan 2017; Jul 2017\\
FCC~184 & $5\times720$ & $6\times600$ & $9\times600$ & Jan 2017; Aug 2017; Sep 2017; Oct 2017; Nov 2017\\
FCC~190 & $5\times720$ & ...          & $9\times600$ & Nov--Dec 2017\\
FCC~193 & $5\times720$ & ...          & $9\times600$ & Nov--Dec 2017\\
FCC~213 & ...          & $9\times600$ & $12\times600$ & Oct 2016\\
FCC~219 & $5\times720$ & ...          & $9\times600$ & Nov 2017\\
FCC~249 & $5\times720$ & ...          & $3\times600$ & Nov 2017\\
FCC~255 & $5\times720$ & ...          & $3\times600$ & Nov 2017\\
FCC~263 & $4\times900$ & ...          & ...             & Jul--Aug 2016; Nov 2017\\
FCC~276 & $5\times720$ &...           & $9\times600$ & Aug 2016; Oct 2016; Jan 2017; Nov 2017\\
FCC~277 & $5\times720$ & ...          & $3\times720$ & Oct 2016; Oct 2017\\
FCC~285 & $4\times900$ & ...          & ...             & Aug 2016; Jan 2017\\
FCC~290 & $4\times900$ & ...          & $4\times900$ & Aug 2016; Jan 2017; Jul 2017\\
FCC~301 & $5\times720$ & ...          & $3\times600$ & Oct 2016; Nov 2017\\
FCC~306 & $4\times900$ & ...          & ...             & Aug 2016; Dec 2016\\
FCC~308 & $4\times900$ & ...          & $4\times900$ & Oct 2016; Jul--Aug 2017\\
FCC~310 & $5\times720$ & ...          & $9\times600$ & Sep 2017; Nov--Dec 2017\\
FCC~312 & $4\times900$ & $4\times900$ & $4\times900$ & Oct 2016; Dec 2016; Jan--Feb 2017; Sep 2017; Nov 2017\\
\noalign{\smallskip}
\hline
\end{tabular}
\tablefoot{(1) Galaxy identification from \citet{Ferguson1989}. (2),
  (3), and (4) Total exposure time for central, middle, and halo
  pointings (see Fig.~\ref{fig:pointings_dss}). (5) Date of the
  observations. The single central pointing of FCC~113, FCC~176,
  FCC~263, and FCC~306 covers also the outskirts of these
  galaxies.\looseness=-2}
\end{table*}
}

\begin{document}

\title{The Fornax3D project: overall goals, galaxy sample, MUSE data
  analysis and initial results}

\titlerunning{The Fornax3D project}
\subtitle{}

\author{M.~Sarzi\inst{1}\thanks{m.sarzi@herts.ac.uk}
\and E.~Iodice\inst{2}
\and L.~Coccato\inst{3}
\and E.~M.~Corsini\inst{4, 5}
\and P.~T.~de~Zeeuw\inst{6, 7}
\and J.~Falc\'on-Barroso\inst{8, 9}
\and D.~A.~Gadotti\inst{3}
\and M.~Lyubenova\inst{3}
\and R.~M.~McDermid\inst{10, 11}
\and G.~van~de~Ven\inst{3}
\and K.~Fahrion\inst{3, 12}
\and A.~Pizzella\inst{4, 5}
\and L.~Zhu\inst{12}
}

\institute{
Centre for Astrophysics Research, University of Hertfordshire, College
Lane, Hatfield AL10 9AB, UK
\and INAF--Osservatorio Astronomico di Capodimonte, via Moiariello 16,
I-80131 Napoli, Italy
\and European Southern Observatory, Karl Schwarzschild Strasse 2,
D-85748 Garching bei Muenchen, Germany
\and Dipartimento di Fisica e Astronomia `G. Galilei', Universit\`a di
Padova, vicolo dell'Osservatorio 3, I-35122 Padova, Italy
\and INAF--Osservatorio Astronomico di Padova, vicolo
dell'Osservatorio 5, I-35122 Padova, Italy
\and Sterrewacht Leiden, Leiden University, Postbus 9513, 2300 RA
Leiden, The Netherlands
\and Max-Planck-Institut fuer extraterrestrische Physik,
Giessenbachstrasse, 85741 Garching bei Muenchen, Germany
\and Instituto de Astrof\'isica de Canarias, Calle V\'ia L\'actea s/n,
E-38200 La Laguna, Spain
\and Departamento de Astrof\'isica, Universidad de La Laguna, Calle
Astrof\'isico Francisco S\'anchez s/n, E-38205 La Laguna, Spain
\and Department of Physics and Astronomy, Macquarie University,
Sydney, NSW 2109, Australia
\and Australian Astronomical Observatory, PO Box 915, Sydney, NSW 1670,
Australia
\and
Max-Planck-Institut fuer Astronomie, Koenigstuhl 17,
69117 Heidelberg, Germany}


\abstract{The Fornax cluster provides a uniquely compact laboratory to
  study the detailed history of early-type galaxies and the role
  played by environment in driving their evolution and their
  transformation from late-type galaxies. Using the superb
  capabilities of the Multi Unit Spectroscopic Explorer on the Very
  Large Telescope, high-quality integral-field spectroscopic data were
  obtained for the inner regions of all the bright ($m_B\leq15$)
  galaxies within the virial radius of Fornax. The stellar haloes of
  early-type galaxies are also covered out to about four effective
  radii. State-of-the-art stellar dynamical and population modelling
  allows to aim in particular at better characterising the disc
  components of fast-rotating early-type galaxies, constraining radial
  variations in the stellar initial-mass functions and measuring the
  stellar age, metallicity, and $\alpha$-element abundance of stellar
  haloes in cluster galaxies. This paper describes the sample
  selection, observations, and overall goals of the survey, and
  provides initial results based on the spectroscopic data, including
  the detailed characterisation of stellar kinematics and populations
  to large radii; decomposition of galaxy components directly via
  their orbital structure; the ability to identify globular clusters
  and planetary nebulae, and derivation of high-quality emission-line
  diagnostics in the presence of complex ionised gas.  }


\maketitle

%

\section{Introduction}
\label{sec:intro}

Integral-field spectroscopy has allowed precise mapping of the stellar
and gas kinematics as well as the stellar-population properties in
thousands of nearby galaxies (see, e.g., the SAURON, ATLAS3D, CALIFA,
ManGA, and SAMI surveys described in \citealt{deZeeuw2002},
\citealt{Cappellari2011}, \citealt{Sanchez2012}, \citealt{Bundy2015},
and \citealt{Croom2012}, respectively). This has led to significant
advances in our understanding of galaxy formation and evolution, in
particular regarding early-type galaxies \citep[ETGs; see,
  e.g.,][]{Cappellari2016}. In particular, classical photometric
studies typically divided ETGs into elliptical and lenticular
morphologies --- classifications that were prone to subjective
uncertainties due to effects of projection and data quality. The
detailed mapping of their stellar kinematics enabled by integral-field
spectroscopy, however, has revealed that ETGs are more physically
divided between a minority of slowly-rotating, roundish, and
predominantly massive galaxies, and a majority of fast-rotating
systems that span a wide range in mass and apparent flattening and for
which there is often evidence for stellar discs \citep{Emsellem2007,
  Emsellem2011, Krajnovic2013}. In fact, fast-rotating galaxies (FRs)
are also intrinsically much flatter than slow rotators (SRs) and
nearly as flat as spiral galaxies \citep{Weijmans2014, Foster2017},
which further suggests that the majority of ETGs could have evolved
from late-type galaxies \citep[LTGs; see also][]{Cappellari2013}.
Furthermore, following on the earliest findings by
\citet{vanDokkum2010} and \citet{Treu2010}, integral-field
spectroscopy studies also contributed to establishing the presence of
systematic variations of the stellar initial mass function (IMF) in
ETGs, both across different objects by means of dynamical models
\citep[e.g.,][]{McDermid2015} or within single galaxies through
stellar-population analysis \citep[e.g.,][]{MartinNavarro2015a}.

Despite this tremendous progress, there are still outstanding
questions regarding the formation and evolution of ETGs. For example,
SRs exhibit a high incidence of kinematically distinct components,
often detected as centrally misaligned `cores'. However, the frequency
of kinematically distinct components is at odds with the fragility of
such central components against merging events \citep{Bois2011}, even
though SRs are generally thought to have formed through this path
\citep{Naab2014}. Understanding the true nature and origin of these
decoupled components remains an open issue. In FRs, the photometric,
kinematic, and stellar-population properties of their embedded disc
components need to be accurately separated from those of the host
spheroidal component in order to further understand the evolutionary
connection between FRs and spiral galaxies. Current studies that
attempt this typically assume a single disc component with an
exponential surface-brightness profile embedded in a spheroid with a
steeper surface brightness profile \citep[e.g.,][]{Johnston2013,
  Coccato2015}. Recent findings from sophisticated orbit-superposition
models indicate that the situation may be more complex, and that ETGs
may also contain a dynamically-warm component resembling the `thick
disc' in spiral galaxies \citep{Zhu2018a}.

In terms of stellar populations, more work is needed to corroborate
measurements of a varying stellar IMF in ETGs. While the initial
tension between global stellar dynamical and population IMF results
\citep{Smith2014} has been resolved \citep{Lyubenova2016}, it still
remains to be established whether this agreement holds in spatially
resolved studies of the IMF. While more and more papers report the
presence of radial IMF gradients indicative of central dwarf-rich
stellar populations \citep[e.g.,][]{vanDokkum2017, LaBarbera2017}, the
effect of radial gradients in the abundance of elements entering the
IMF-sensitive absorption-line features \citep[such as
  sodium,][]{Zieleniewski2015, McConnell2016} still needs to be
established.

Moving outward in radius, to date, most efforts to directly map the
stellar-population properties of galactic haloes reach only out to
2.5--4 effective radii $R_{\rm e}$, and with sparse spatial sampling
\citep[][]{Pastorello2014, Arnold2014, Greene2015, Boardman2017,
  Bellstedt2018}. Simulations indicate that the strongest signatures
of past mergers should be more evident further out in the galaxy
outskirts where phase-mixing becomes progressively more inefficient
\citep[e.g.,][]{Hirschmann2015}. Indeed, deep broad-band imaging often
unveils rich structure in the outer parts of ETGs indicative of
major mergers \citep{Duc2015}. Globular clusters have been used to
probe these outer regions \citep[e.g., the SLUGGS
  survey,][]{Brodie2014}, however the relationship between the dynamics
and metallicity distribution of globular clusters with their host
galaxy can be difficult to ascertain given the very different radial
ranges typically probed observationally, leading to conflicting
inferences about halo properties compared to other tracers
\citep[e.g.,][]{Alabi2017}.

With its extended spectral range, fine spatial sampling, large field
of view, and superb throughput, the Multi Unit Spectroscopic Explorer
(MUSE) integral-field spectrograph \citep{Bacon2010} is a unique
instrument to address these (and related) outstanding issues. For
instance, the early work of \citet{Emsellem2014} demonstrates the
power of this instrument for unveiling previously unknown
kinematically-distinct cores in SRs. The study of
\citet{Krajnovic2015} illustrates how well the structure of the inner
regions can be recovered when Schwarzschild orbit-superposition models
are based on high-quality MUSE kinematic data.  Similarly, precise
measurements of higher moments of the stellar line-of-sight velocity
distribution out to several effective galactic radii \citep[such as
  those presented by][]{Guerou2016} will significantly extend the
measurement of the large-scale mass profiles of nearby galaxies
\citep{Poci2017,Bellstedt2018} while simultaneously providing their
orbital distribution \citep[see][]{Zhu2018a}.

The MUSE spectral range also covers several weak absorption-line
features that are sensitive to the fraction of low-mass stars and the
abundance of various elements (see, e.g., the work of
\citealt{Spiniello2014} and \citealt{Conroy2014} on Sloan Digital Sky
Survey data), which makes it possible to trace IMF gradients with
great accuracy while also controlling for radial variation of element
abundances \citep[see, e.g.,][]{Mentz2016, Sarzi2017}. Even more
exciting in this context, is the prospect of measuring stellar
mass-to-light ratio, $M/L$, gradients from dynamical models based on
MUSE data \citep{Oldham2018}, and use those to constrain IMF gradients
from stellar-population models.

The outstanding continuum signal-to-noise ratio, $S/N$, attainable
with MUSE permits high-quality measurements of the stellar population
and star-formation history diagnostics within modest exposure times,
whilst retaining the fine spatial resolution required to analyse
galaxy substructures and recover meaningful maps of star-formation or
metal-enrichment histories \citep[see, e.g.,][]{Guerou2016}. Further
characterisation of the stellar population properties of the embedded
disc components of FRs may become possible if the results of
orbit-superposition models can be combined with spectral-analysis
approaches such as those of \citet{Johnston2013} or
\citet{Coccato2015}, thus providing physically-motivated priors on the
relative light contribution and kinematics of dynamically-distinct
hot, warm, and cold components.\looseness=-2

The exquisite sensitivity and high stability of MUSE data allow
effective stacking of neighbouring spectra with minimal systematic
errors, raising the potential to constrain the stellar-population
properties of galactic haloes out to several $R_{\rm e}$, following
the example of \citet{Weijmans2009}. Moreover, MUSE can measure the
radial velocities as well as the chemical composition of globular
clusters (GCs) in nearby galaxies, providing additional constraints on
halo properties. The GCs not only provide discrete tracers of the
underlying gravitational potential to constrain the total mass
distribution of their host galaxy \citep[e.g.,][]{Zhu2018a}, but their
orbits and chemistry also contain a fossil record of where they were
born. In particular, while the stars of merging satellite galaxies
quickly disperse, their surviving GCs still provide a way to uncover
the hierarchical build-up of the host galaxy. Similarly, following on
the earliest steps done with SAURON observations in our Galactic
neighbours M32 and M31 \citep{Sarzi2011,Pastorello2013}, MUSE
spectroscopic data allow a systematic investigation of the properties
of the planetary nebulae (PNe) in the bright optical regions of ETGs
that so far have remained largely unexplored.

The Fornax cluster of galaxies offers an ideal environment to study
the archaeological record of the formation of ETGs, for several
reasons. It is the second closest galaxy cluster, at a distance of
only 20.9 ($\pm0.9$) Mpc \citep{Blakeslee2009}, which allows a
detailed view of the galaxies and the collection of high
signal-to-noise data. At a declination of about $-35^\circ\,30'$, the
Fornax cluster is perfectly located for observations with telescopes
at ESO's La Silla--Paranal Observatory. HST images \citep{Jordan2007}
and VST deep multi-band imaging \citep{Iodice2016, Venhola2017} are
available for the ETGs in this cluster, which is essential for
stellar-dynamical modelling of their central regions and for the
stellar-population analysis of their haloes. The cluster has a modest
mass (about $7 \times 10^{13}$ M$_{\odot}$) compared to Virgo or Coma,
which are 10 times more massive, making Fornax an important lower-mass
counterpoint to these well-studied, nearby clusters. It is also
compact and dynamically relaxed, and therefore is more representative
of the clusters and groups in which most ETGs tend to live (see, e.g.,
the cluster mass function of \citealt{Bahcall1993}). ETGs in clusters
are also unlikely to suffer (further) major interactions
\citep{Mihos2005}, and gas accretion and star formation are likewise
suppressed \citep{Davis2011, McDermid2015}, in particular in the
$10^{13}-10^{14}$ M$_\odot$ range \citep{Catinella2013,
  Odekon2016}. Therefore, the Fornax cluster gives a clean view into
the assembly history of ETGs since they joined the cluster, with
minimal significant external influences on their dynamical or chemical
evolution. Finally, as the Fornax cluster appears particularly rich in
FRs \citep{Scott2014}, it provides an ideal laboratory for testing the
role of galactic environment in quenching star formation and turning
spirals into ETGs.

These considerations motivated the initiation of the Fornax3D project,
a timely and comprehensive study of the galaxies inside the virial
radius of the Fornax cluster based on observations with MUSE. The MUSE
data provide a rich and detailed picture of the cluster's principal
central members, allowing extensive study of their resolved and
unresolved stellar and ionised gas properties using state-of-the-art
modelling techniques, and complementing the extensive high-quality
data existing for this nearby cluster. This paper presents an overview
of the project, and highlights some initial results that demonstrate
the quality, legacy value, and exciting potential of the Fornax3D data
set. The design of the F3D survey is described in
Sect.~\ref{sec:plan}. The observations and data reduction are
discussed in Sect.~\ref{sec:obs}. The data analysis is summarised in
Sect.~\ref{sec:analysis} and quality of the data is assessed in
Sect.~\ref{sec:quality}. A number of initial results are given in
Sect.~\ref{sec:results} and concluding remarks follow in
Sect.~\ref{sec:remarks}.

\section{Survey design}
\label{sec:plan}

\begin{figure}[t!]
\centering
\includegraphics[width=\hsize]{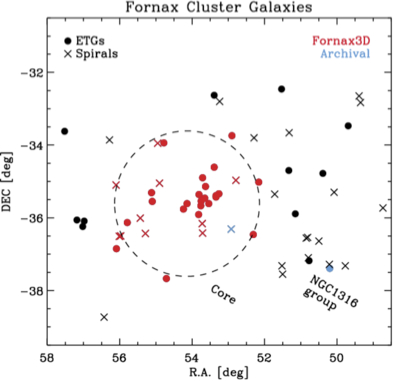}
\caption{Location on the sky of the ETGs (circles) and LTGs (crosses)
  with $m_B<15$ mag in the Fornax cluster. The dashed circle marks the
  virial radius of the cluster and corresponds to about $0.7$ Mpc, as
  determined by \citet{Drinkwater2001}. The red symbols show F3D
  galaxies while the blue symbols correspond to NGC~1316 (filled
  circle) and NGC~1365 (cross) for which MUSE data are already available in the ESO
  Science Archive Facility.}
\label{fig:fornax}
\end{figure}

The F3D sample consists of 33 galaxies, 23 of which are ETGs
(Table~\ref{tab:sample}). They were selected from the catalogue of the
Fornax cluster members by \citet{Ferguson1989}, as the galaxies
brighter than $m_B=15$ mag within or close to the virial radius
\citep[$R_{\rm vir} \sim 0.7$ Mpc,][]{Drinkwater2001}.  Fainter
objects are all classified as dwarf galaxies and their study is beyond
the scope of this project. Fig.~\ref{fig:fornax} shows the location of
the F3D galaxies in the Fornax cluster. There are a few LTGs inside
the virial radius of the cluster. They were included in the F3D sample
to provide a high-density counter-part to other MUSE studies of spiral
galaxies in the field \citep[e.g.,][]{Gadotti2015} and to further help
to understand the link between LTGs and FRs \citep[as done in the
  framework of the CALIFA survey by][]{FalconBarroso2016}.

To reach the scientific objectives of the project, the centre and
outskirts of all the sample ETGs out to the stellar halo \citep[i.e.,
  where the surface brightness is $\mu_B \geq 25$ mag~arcsec$^{-2}$,
  see][]{Janowiecki2010, Spavone2017, Iodice2016, Iodice2017a} were
mapped with MUSE. For the majority of these galaxies, the extended
surface photometry in the $B$ band provided by \citet{Caon1994} was
used to define the halo pointings. They were chosen to both overlap
the central pointing and cover a portion of the galaxy isophote at
$\mu_B = 25$ mag~arcsec$^{-2}$. The semi-major axis length $R_{25}$ of
the best-fitting ellipse to this isophote is listed in
Table~\ref{tab:sample}. For the remaining galaxies, the size and
orientation of the isophote at $\mu_B = 25$ mag~arcsec$^{-2}$ were
taken from \citet{RC3}.  A single MUSE pointing was sufficient to
cover both the central regions and galaxy outskirts of the ETGs with a
major-axis diameter $2R_{25} < 1$ arcmin. For the larger ETGs, two or
even three MUSE pointings were needed to map the whole galaxy. In the
latter case, the middle pointing targeted the transition region
between the central and halo pointings. For most of the F3D LTGs, only
the central regions were observed.  The exceptions were the most
extended LTGs of the sample (FCC~290, FCC~308, and FCC~312), which
also merited two or three pointings. The MUSE pointings of all the F3D
galaxies are listed in Table~\ref{tab:log} and shown in
Fig.~\ref{fig:pointings_dss}.

\begin{figure}[t!]
\centering
\includegraphics[width=4.5cm, angle=90]{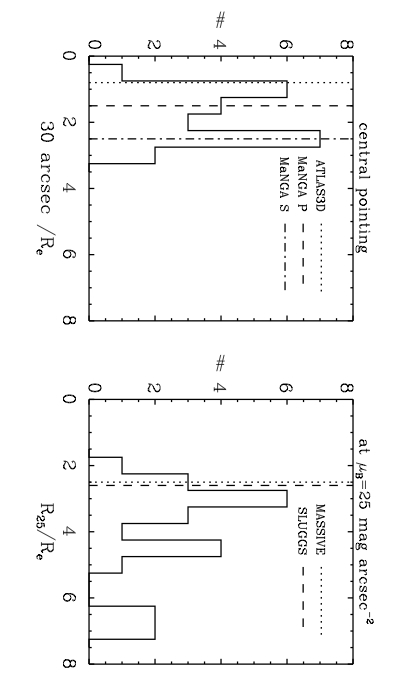}
\caption{{\it Left panel:\/} Spatial coverage of the MUSE central
  pointings of the F3D ETGs in units of $30\,{\rm arcsec}/R_{\rm e}$
  (histogram) compared to the ATLAS3D \citep[][dotted
    line]{Cappellari2011} and MaNGA \citep[][primary sample: dashed
    line, secondary sample: dash-dotted line]{Bundy2015} surveys. {\it
    Right panel:\/} Spatial coverage of the MUSE halo pointings of the
  F3D ETGs down to $\mu_B = 25$~mag arcsec$^{-2}$ in units of
  $R_{25}/R_{\rm e}$ (histogram) compared to the MASSIVE
  \citep[][dotted line]{Greene2015} and SLUGGS \citep[][dashed
    line]{Foster2016} surveys.}
\label{fig:coverage}
\end{figure}

High spectral quality in the central regions (within $2R_{\rm e}$) was
needed to extract high-order moments of the stellar line-of-sight
velocity distribution and construct Schwarzschild orbit-superposition
models to study the kinematically decoupled cores in SRs and the discs
of FRs, separate the disc and bulge stellar populations in FRs,
measure radial variations in the IMF, map the chemical abundance of
various elements, and carry out a census of the population of the
central GCs and PNe. Following \citet{Krajnovic2015}, this required a
target $S/N=100$ per spectral pixel at 5500 \AA\ which was reached in
one hour of on-source integration time for most of the F3D galaxies
with a modest spatial binning even close to the edge of MUSE field of
view. Indeed, except for a few compact objects, $\mu_B \sim 22$
mag~arcsec$^{-2}$ was measured at 30 arcsec from the galaxy centre
(Table~\ref{tab:sample}). At this surface brightness level, the target
$S/N$ was reached within an area of $2.8 \times 2.8$ arcsec$^{2}$
(corresponding to $14 \times 14$ spaxels). This spectral quality
should also allow to measure IMF gradients based on line-strength
indices \citep[e.g.,][]{Sarzi2017} and to constrain chemical
abundances using direct spectral fitting, for which a $S/N \sim 40$
per spectral pixel should suffice \citep{Choi2014}.

\placetabone

The regions of the stellar halo down to $\mu_B \geq 25$
mag~arcsec$^{-2}$ were reached in the outskirts of ETGs (out to $R
\geq 4R_{\rm e}$). At such low surface-brightness levels, MUSE
observations have already proved to deliver good-quality spectra
\citep{Iodice2015}. An on-source integration time of 1.5 hours within
reasonable-sized spatial bins of $6\times6$ arcsec$^{-2}$
(corresponding to $30 \times 30$ spaxels) allowed to get a target
$S/N=25$ per spectral pixel at $5500$ \AA . This was needed to
constrain the stellar age, metallicity, and $\alpha$-element abundance
\citep{Choi2014}. Thus, in general, the total on-source integration
time for each ETG was 2.5 hours to cover both the central and outskirt
regions. For smaller objects ($2R_{25} <1$ arcmin), where the central
pointing covers also the galaxy outskirts, the integration time was
reduced to 1.5 hours. MUSE observations of the central regions of
NGC~1399, the brightest member of the Fornax cluster, were already
available in the ESO Science Archive Facility (Prop. Id.
094.B-0903(A), P.I. S. Zieleniewski). Therefore, new data were
acquired in the transition and outskirt regions with a middle and halo
pointing. The spatial coverage of central and halo pointings for F3D
ETGs with a comparison with other integral-field spectroscopic surveys
is shown in Fig.~\ref{fig:coverage}.

The central regions of the less extended LTGs were observed with one
hour of on-source integration (Table~\ref{tab:log}) in order to reach
the same limiting magnitude as for the halo regions in ETGs.  The
bright barred spiral galaxy NGC~1365, in the southwest part of the
Fornax cluster (Fig.~\ref{fig:fornax}), was excluded from the F3D
sample since suitable MUSE observations were present in the ESO
Science Archive Facility (Prop. Id. 094.B-0321(A), P.I. A. Marconi),
and the galaxy is part of the sample of the TIMER survey
\citep{Gadotti2018}, which already produced high-level data products
from the archival data.

\section{Observations and data reduction}
\label{sec:obs}

The integral-field spectroscopic observations of the F3D sample
galaxies were carried out with MUSE mounted on the Yepun Unit
Telescope 4 at the ESO Very Large Telescope in Chile. MUSE was
configured in Wide Field Mode \citep{Bacon2010} that ensured a $1
\times 1$ arcmin$^2$ field coverage with $0.2 \times 0.2$ arcsec$^2$
spatial sampling and a wavelength range of 4650--9300 \AA\ with
nominal spectral resolution of 2.5 \AA\ (FWHM) at 7000 \AA\ and
spectral sampling of 1.25 \AA\ pixel$^{-1}$.

Data were taken in service mode between July 2016 and December 2017
(Table~\ref{tab:log}). Typical integration times of on-source
exposures were 12 minutes for the central pointings and 10 minutes for
the middle and halo pointings. The on-source exposures were dithered
by a few arcseconds and rotated by 90 degrees in order to average the
spatial signature of the 24 integral-field units on the field of
view. For the majority of the targets, dedicated sky exposures of
three minutes were scheduled within each observing block, immediately
before or after of each on-source exposure.

The data reduction was performed with the MUSE pipeline version 1.6.2
\citep{Weilbacher2012, Weilbacher2016} under the ESOREFLEX environment
\citep{Freudling2013}. The reduction cascade included bias and
overscan subtraction, flat fielding to correct the pixel-to-pixel
response variation, wavelength calibration, determination of the line
spread function, illumination correction with twilight flats to
account for large-scale variation of the illumination of the
detectors, and illumination correction with lamp flats to correct for
the edge effects between the integral-field units. For each galaxy,
the twilight exposures were combined following the same observing
pattern of the target observations, producing a master twilight
datacube that was used to determine the effective spectral resolution
and its variation across the field of view.

The sky subtraction was done by fitting and subtracting a sky model
spectrum on each spaxel of the field of view. The sky model was either
evaluated on dedicated sky observations or on the edges of the
exposure for the smallest targets, where the contribution of the
galaxy was negligible. The flux calibration and the first-order
correction of the atmospheric telluric features were obtained using
the spectro-photometric standard star observed at twilight.

Each exposure produced a fully reduced `pixel table'. For each
galaxy, the pixel tables were aligned using reference sources and then
combined in an single datacube. For each individual galaxy, a datacube
with the sky residuals was also produced. It was obtained by combining
the sky-subtracted pixel tables of the offset sky exposures in a
single datacube following the same rotation pattern as the target
observations.

An additional cleaning of the residual sky contamination was achieved
with the Zurich Atmospheric Purge algorithm \citep[ZAP,][]{Soto2016},
which exploits principal component analysis to characterise and
subtract the sky residuals from the observations. The principal
components of the sky residuals were evaluated on the sky-residual
datacubes or at the edges of the sky-subtracted target datacube,
depending on the spatial extension of the galaxy. The algorithm was
run within the ESOREFLEX environment, using the dedicated work flow
distributed with the MUSE pipeline. For a number of targets, the
results were compared by running ZAP on the individual exposures
before the creation of the final datacube, and on the final datacube
itself. No significant differences were found. Therefore, the adopted
strategy was to clean the final combined datacube and not the
individual exposures.

\placetabtwo

\section{Data analysis}
\label{sec:analysis}

The data analysis was built on tools and techniques that were
developed for the SAURON, ATLAS3D, and CALIFA surveys.

The Voronoi binning scheme by \citet{Cappellari2003} was used to
spatially bin the data to a minimum $S/N$ per pixel to ensure a
reliable extraction of the relevant parameters of the stellar and
ionised-gas kinematics and stellar populations. The stellar kinematics
was computed using the Penalised Pixel-Fitting code
\citep[pPXF,][]{Cappellari2004, Cappellari2017}. Emission-line
morphologies and kinematics were derived using the Gas and Absorption
Line Fitting code \citep[GandALF,][]{Sarzi2006,
  FalconBarroso2006}. The line-strength indices for the stellar
population analysis were extracted as well. In the remainder of this
section, the specific details about how all these methods were applied
to the F3D data are provided.

The Voronoi binning process started by selecting spaxels with a
minimum $S/N=3$ (median value per pixel) as estimated using the
pipeline variance cube. This was necessary to avoid poor quality
spaxels which could introduce undesired systematic effects in the data
at low surface-brightness regimes, that are not straightforward to
account for (e.g., affecting the accuracy of the sky subtraction). In
order to produce maps that match the surface photometry of the
galaxies, all spaxels within the isophote level where $\langle S/N
\rangle \sim 3$ were selected. Given that the survey has a wide range
of scientific goals with different requirements, Voronoi binned maps
were produced to $S/N=40$ and 100.

The combination of individual dithered observations introduced spatial
correlations between neighbouring spatial and spectral pixels in the
final combined data. This mainly affected the variance spectra
resulting from the propagation of uncertainties at every step of the
full data reduction process. While in principle, this effect could be
accounted for by computing the covariance matrix for this combination
step, the sheer volume of the data made this approach unfeasible and
computationally very expensive. Instead, the more pragmatic solution
followed by the CALIFA survey \citep{GarciaBenito2015} was adopted of
empirically characterising the level of correlation and correcting for
it during the Voronoi binning process. However, as discussed in
Sect.~\ref{sec:quality}, the level of covariance in the data is
relatively small compared to that in the CALIFA survey.

The extraction of the stellar kinematics in the Voronoi binned spectra
is restricted to the 4750--5500 \AA\ wavelength range. Use of the full
wavelength coverage of the MUSE datacube did not alter the results in
any significant manner. The advantages were a major speed up in
computation time and avoiding to correct the stellar templates for
spectral resolution changes along wavelength. In such a short spectral
range, only a low-order additive polynomial (${\rm degree} = 8$) was
necessary to compensate for differences in calibration between the F3D 
data and templates. The MIUSCAT models \citep{Vazdekis2012} were
adopted as stellar templates for the extraction of the stellar
kinematics. A subset of 65 models was selected that uniformly sample
age (from 0.1 to 15.8 Gyr) and metallicity (from $-2.32$ to 0.22
dex). This subset was a good representation of the entire MIUSCAT
library and very few differences were found between the two sets.

By contrast with the stellar kinematics, for which the applied spatial
binning was based on the $S/N$ estimated from the stellar continuum,
the analysis of the emission-line properties was done on a
spaxel-by-spaxel level, in order to capture the structure in the
emission-line distribution at the highest spatial resolution provided
by the data. GandALF was used in the wavelength range between
4800--6800 \AA, using only a two-component reddening correction to
adjust for the stellar continuum shape and for the observed Balmer
decrement. Similar to the standard setup used in Sloan Digital Sky
Survey data in the value-added catalogue of \citet{Oh2011}, a table
was created with fitted emission-line species and relative
dependencies. The stellar kinematics assumed for each spaxel is that
of the Voronoi bin each spaxel belongs to. Given the large differences
in wavelength of the different emission lines, spectral resolution
variations are accounted for during the fitting process. Note that at
this point that {\it spatial\/} variations of the spectral resolution
within the field of view of MUSE were not taken into account.

The study of the stellar population requires high $S/N$. It is for
that reason that the line-strength analysis was based on the Voronoi
binned data. Since emission-line contamination of stellar absorption
features is an issue, the procedures summarised above were repeated
with GandALF for the Voronoi binned spectra.  Nebular emission as a
whole was conservatively considered as detected when the
amplitude-over-noise ratio was above 3 for all major lines (e.g.,
H$\alpha$, H$\beta$, [\ion{N}{ii}], and [\ion{O}{iii}]).  Emission was
removed from the spectra when, based on the GandALF uncertainties
there was a $68\%$ probability that the no-emission hypothesis could
be rejected.  Line-strength indices were computed in the LIS system
\citep{Vazdekis2010}. This approach minimised some of the major
intrinsic uncertainties associated with the popular Lick/IDS system
\citep{Gonzalez1993}. This line-strength index system is based on
flux-calibrated spectra with a constant resolution as a function of
wavelength providing three different resolutions, namely 5, 8.4, and
14 \AA\ (FWHM). The resolution of 8.4 \AA, which is appropriate for
studying low- to intermediate-mass galaxies, was usually sufficient
for F3D scientific aims.

In addition to line-strength indices, the use of regularised solutions
in pPXF was also explored to derive star formation histories across
the entire MUSE maps. This has the potential to reveal the formation
sequence of the different components in galaxies and therefore provide
a much more complete evolutionary picture. First results following
this approach are presented in \citet{Pinna2018} to study the origin
of the thick disc component in FCC~170.

The instrumental spectral resolution can be determined by the MUSE
pipeline by recovering the full shape of the line spread function
using a series of dedicated arc lamp observations. The instrumental
spectral resolution varies with wavelength, integral-field unit, and
slice. The galaxy observations were composed of exposures taken at
varying orthogonal position angles, thus the combined spectra are
influenced by each of these effects. The following strategy was
adopted to measure the resulting line spread function. For each
galaxy, $N$ twilight datacubes were combined using the same
observational pattern (including rotations and offsets) as the $N$
galaxy exposures. The resulting effective instrumental broadening
function was derived from the combined twilight cubes by applying pPXF
with a solar-spectrum template. The process was divided in a number of
wavelength intervals and for each spaxel in order to measure the
variation of the fitting parameters with wavelength and position over
the field of view. On average, the instrumental spectral resolution
was ${\rm FWHM_{\rm inst}} \sim 2.8$ \AA\ with little variation ($<
0.2$ \AA ) with wavelength and position over the field of view. This
showed that the effect of having combined exposures taken at different
rotations (i.e., of having combined the signal from different
integral-field units) is an increase of the mean instrumental FWHM and
a homogenisation over the field of view and wavelength. A modest
spatial dependence of the FWHM remains, however. The twilight maps can
be used to correct for this where needed (e.g., when the velocity
dispersion is close to or less than the instrumental resolution).

\section{Data quality}
\label{sec:quality}

In this section the MUSE data obtained for FCC~167 are analysed to
assess the quality of the data, with attention to {\it i)\/} the $S/N$
per spaxel down to surface brightness of $\mu_B = 25$
mag~arcsec$^{-2}$, which includes  a quality check of the subtraction
of the sky emission; {\it ii)\/} the correction for the atmospheric
telluric absorption which is key to constraining the low-mass end of
the IMF from a spectral analysis; {\it iii)\/} the quality of the
relative flux calibration, which is related to the ability to measure
IMF-sensitive absorption-line index (e.g., aTiO, TiO$_1$, and
TiO$_2$); {\it iv)\/} the quality of the absolute flux calibration,
which is important for the extraction of reliable emission-line flux
values and thus derive, for instance the star-formation rate or
luminosity of PNe.

\begin{figure}[t!]
\centering
\includegraphics[width=\hsize]{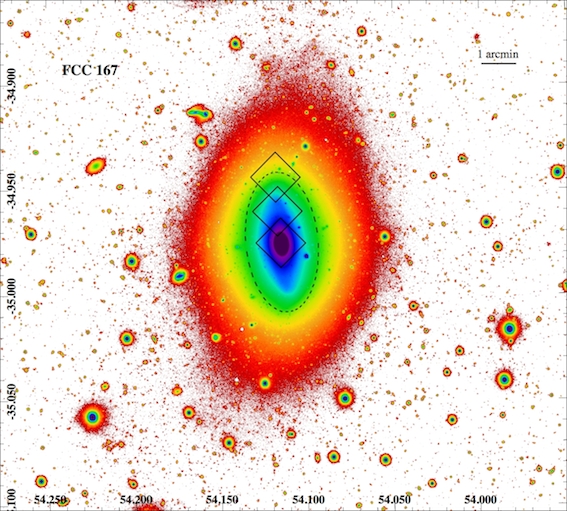}
\caption{$r$-band image of FCC~167 from the Fornax Deep Survey with
  VST \citep{Iodice2016, Venhola2017}. The $1 \times 1$ arcmin$^2$
  MUSE pointings are shown in black. The black dashed ellipse
  corresponds to the isophote at $\mu_B=25$ mag~arcsec$^{-2}$, while
  the right ascension and declination (J2000.0) are given in degrees
  on the horizontal and vertical axes of the field of view,
  respectively.\looseness=-2}
\label{fig:pointings_fds}
\end{figure}

\begin{figure}[t!]
\centering
\includegraphics[width=\hsize]{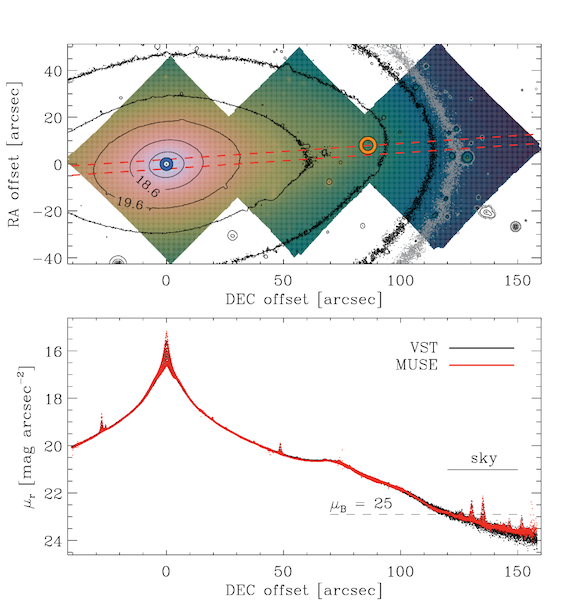}
\caption{{\it Top panel:\/} MUSE reconstructed image of the combined
  mosaic of FCC~167 compared with isophotes of the $r$-band VST image
  shown in Fig.~\ref{fig:pointings_fds}. The black isophotes cover the
  range between $\mu_r = 16.6$ and 22.6 mag arcsec$^{-2}$ and are
  spaced by 1 mag arcsec$^{-2}$. The grey isophote corresponds to
  $\mu_B = 25$ mag arcsec$^{-2}$.  The blue and orange circles
  indicate, respectively, the position of the central and outer
  apertures selected for testing the spectral quality of the F3D data (see
  Fig.~\ref{fig:quality_spectroscopy}).  The red dashed lines mark the
  2-arcsec wide aperture along the galaxy major axis ($\rm P.A. = 3.9$
  degrees) where the surface-brightness radial profile plotted in the
  bottom panel was extracted.  {\it Bottom panel:\/}
  Surface-brightness radial profile along the major axis of FCC~167
  from the MUSE reconstructed image (red points) and $r$-band VST
  image (black points). The sky level of MUSE (horizontal continuous
  line) and $\mu_B=25$ mag~arcsec$^{-2}$ (horizontal dashed line) are
  shown for comparison.}
\label{fig:quality_photometry}
\end{figure}

FCC~167 is an S0/a galaxy on the northeast side of the Fornax cluster,
located at about $0.6$~degrees (corresponding to $222$~kpc) from
NGC~1399 and at a distance of 21.2 Mpc from surface-brightness
fluctuations \citep{Blakeslee2009} (see Table~\ref{tab:sample}). The
three pointings for this object cover the central and middle regions
and the outskirts (Table~\ref{tab:log}).  Fig.~\ref{fig:pointings_fds}
shows the MUSE pointings on an $r$-band image obtained with the Fornax
Deep Survey \citep{Iodice2016, Venhola2017} and
Fig.~\ref{fig:quality_photometry} compares the MUSE reconstructed
image for the combined mosaic with the isophotes in the $r$ band
derived from the deep VST image shown in
Fig.~\ref{fig:pointings_fds}. The top panel of
Fig.~\ref{fig:quality_photometry} demonstrates that the individual
pointings were correctly aligned before combining them, whereas the
lower panel shows more quantitatively by means of a major-axis cut how
well the surface-brightness radial profile of the MUSE reconstructed
image, after arbitrarily rescaling, follows that of the VST
image. This comparison also illustrates that the overall level of the
sky subtraction has been determined correctly, in particular down to
the target surface-brightness level of $\mu_B = 25$ mag~arcsec$^{-2}$
and out to $R_{25}/R_{\rm e} \sim 3.5$.

The HST images obtained by \citet{Jordan2007} with the Advanced Camera
for Survey (ACS) in the F850LP passband, which covers the red end of
the MUSE spectra (above $8000$ \AA), allow a further check of the
absolute flux calibration. Including a small systematic correction to
account for the fact that this passband extends slightly beyond the
MUSE data, the average flux density of the spectra within the F850LP
passband is within $10\%$ of the flux density measured by the HST
images.

\begin{figure}[t!]
\centering
\includegraphics[width=\hsize]{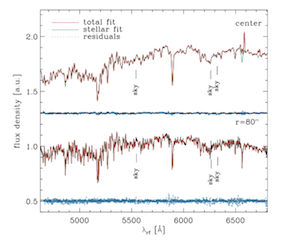}
\caption{GandALF fits to the central ({\it top panel\/}) and outer
  aperture ({\it bottom panel\/}) spectra extracted from the MUSE data
  of FCC~167 (see Fig.~\ref{fig:quality_photometry}). To allow a more
  direct comparison between these two spectra (black lines) and their
  respective GandALF fits (green and red lines), both data and models
  were normalised and rescaled. The residuals obtained by subtracting
  the model from the spectrum are also shown (blue points). These
  models used the entire MILES stellar library as stellar templates to
  achieve the best possible fit. At the same time, to ensure a
  physically-motivated model only reddening by dust was used to adjust
  the templates to the observed shape of the stellar continuum. The
  position of the most prominent [\ion{O}{i}] sky emission lines is
  indicated.}
\label{fig:quality_spectroscopy}
\end{figure}

The quality of the spectroscopic data is best revealed through the
comparison with a physically-motivated fit to both the stellar
continuum and nebular emission. Fig.~\ref{fig:quality_spectroscopy}
shows the best GandALF fit to the two aperture spectra extracted at
the locations indicated in Fig.~\ref{fig:quality_photometry}: one
centred on the nucleus of FCC~167 and the other at about 80 arcsec to
the north, where the surface brightness of the galaxy is comparable to
the sky background.  Only dust reddening was included to correct the
continuum shape, as opposed to resorting to a high-order
multiplicative polynomial correction which may adjust for possible
shortcomings in the data (such as an imperfect relative flux
calibration). For both aperture spectra,
Fig.~\ref{fig:quality_spectroscopy} shows that the quality of the data
is excellent, leading to residuals with little or no significant
structure, except for {\it a)} systematic fluctuations introduced by
template mismatch in the case of the GandALF fit to the very high
$S/N$ nuclear spectrum and {\it b)} some sky-subtraction residuals in
the case of the outer spectrum, due to bands of weak sky
emission-lines (between 5900 and 6650 \AA). The residuals do not show
the presence of low-frequency fluctuations, thus indicating that the
relative flux calibration is under control.

To further evaluate the spectral quality, the
signal-to-fit-residual-noise ratio, $S/rN$, was computed in the
Voronoi-binned spectra and based on the results of the adopted
standard pPXF fitting procedure. This $S/rN$ is a more conservative
measurement of the quality of the MUSE spectra than the formal $S/N$
values based on the variance spectra returned by the data reduction
and their formal propagation during the spatial binning process. This
is because the fit residuals can also capture systematic errors in the
data reduction and in particular spatial covariance introduced during
the combination of individual dithered observations.

\begin{figure}[t!]
\centering
\includegraphics[width=\hsize]{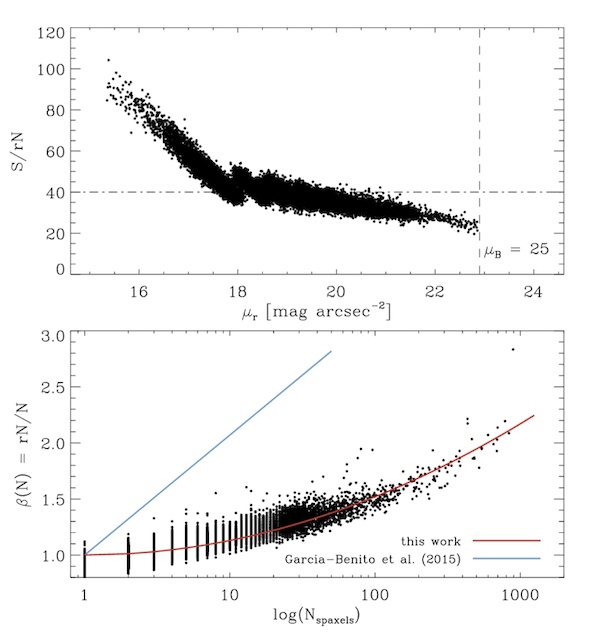}
\caption{{\it Top panel:} Quality of the Voronoi-binned spectra for a
  target $S/N=40$ as a function of $r$-band surface brightness in MUSE.jpg
  data of FCC~167. The $S/rN$ ratio between the median value of the
  flux and the noise level in the fit residuals is used as a more
  conservative measure of the spectral quality.  {\it Lower panel:}
  Level of spatial correlation in the MUSE data of FCC~167. The
  covariance ratio $\beta = rN/N$ between the fit residual noise and
  the formally propagated noise increases as function of bin size (or
  number of spaxels) in each Voronoi bin, although much less than in
  the CALIFA data (blue line). The $\beta$ trend is well described by
  a quadratic form in $\log{N_{\rm spaxels}}$ (red line).}
\label{fig:SN}
\end{figure}

Fig.~\ref{fig:SN} (top panel) shows the values of the $S/rN$ ratio as
a function of the mean $r$-band surface brightness inside the Voronoi
bins obtained imposing a formal $S/N=40$ {\it before\/} further
accounting for spatial correlations, as described in
\citet{GarciaBenito2015}. This illustrates the negative impact of
spatial correlations, as the target $S/N=40$ is not met at first in
the outskirts of FCC~167 where the data were combined in larger
bins. However, the depth of the exposure was sufficient to reach the
target $S/N=25$ per spaxel and resolution element around 5500
\AA\ within the adopted fitting range at a surface brightness of
$\mu_B = 25$ mag~arcsec$^{-2}$ (corresponding $\mu_r \sim 25$
mag~arcsec$^{-2}$ for old stellar populations).

\begin{figure}[t!]
\centering
\includegraphics[width=\hsize]{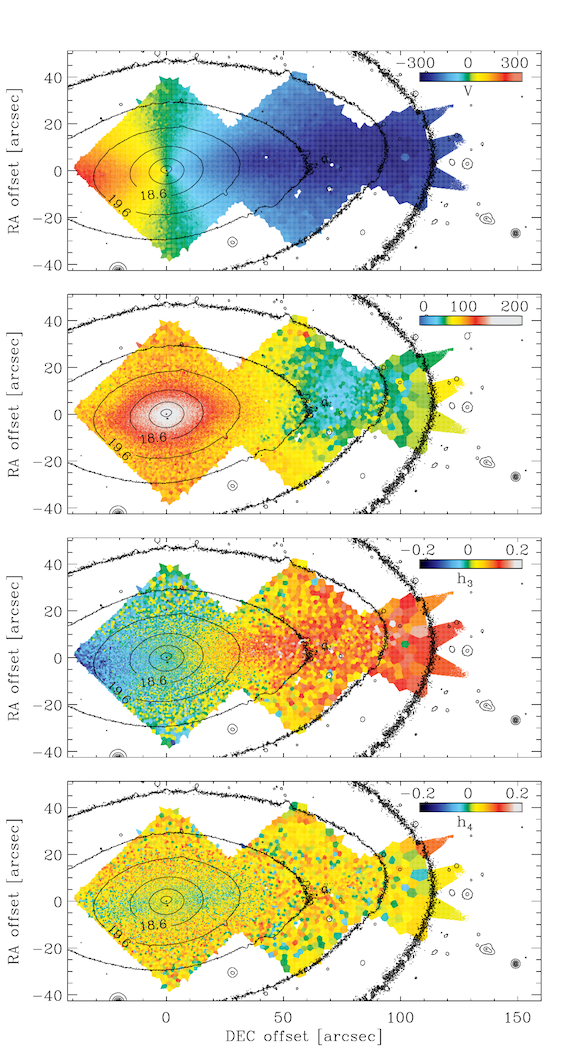}
\caption{Stellar kinematic maps of the mean velocity $v$, velocity
  dispersion $\sigma$, and higher-order moments $h_3$ and $h_4$ of the
  line-of-sight velocity distribution for FCC~167 derived from the
  $S/N=40$ Voronoi-binned MUSE data with the $r$-band isophotes (black
  lines) shown in Fig.~\ref{fig:quality_photometry}.}
\label{fig:FCC167_maps}
\end{figure}

To better quantify the impact of spatial correlations,
Fig.\ref{fig:SN} (lower panel) shows the value of correlation ratio
$\beta = rN/N = (S/N)/(S/rN)$ per Voronoi bin as a function of the
number of spaxels $N_{\rm spaxels}$ between the noise level in the fit
residuals $rN$ and formally propagated noise $N$. As found in
\citet{GarciaBenito2015}, $\beta$ is a linear function of $\log(N_{\rm
  spaxels})$ for relatively small bin sizes, although the spatial
correlation in the MUSE cubes is much smaller than those in the CALIFA
data. For larger bins in low-surface brightness and progressively more
sky-dominated regions, $\beta(N)$ increases faster as a function of bin
size, so that the overall quadratic form $\beta(N) = 1+a(\log{N_{\rm
    spaxels}})^b$ provides a better way to characterise the spatial
correlations. Using this parametrisation to account for covariances
during the Voronoi binning brings the quality of the outer binned
spectra for FCC~167 back to the $S/N=40$ target (see
Fig.~\ref{fig:FCC167_maps} for the resulting kinematics).

The use of the MOLECFIT \citep{Smette2015} and SKYCORR
\citep{Noll2014} procedures provides a better correction for
atmospheric telluric absorption than the one based on the use of
telluric standards (Sect.~\ref{sec:analysis}).  Most of the
improvement is in the quality of the sky-subtraction, and this will be
applied on an individual spectrum basis depending on the science
goals.

\section{Illustrative initial results}
\label{sec:results}

This section presents some preliminary results obtained for the main
science topics listed in Sect.~\ref{sec:intro}. A detailed analysis
and discussion will be presented in forthcoming papers.

\subsection{Embedded discs in fast rotating ETGs}
\label{sec:FRs}

\begin{figure*}[t!]
\centering
\includegraphics[width=\hsize]{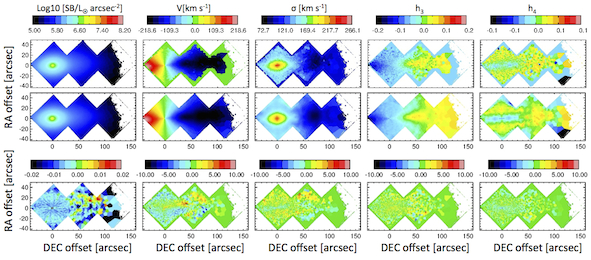}
\caption{Best-fitting Schwarzschild orbit-based dynamical model of
  FCC~167. The {\it top and middle panels\/} show the observed and
  best-fitting model of the surface brightness SB, mean velocity $v$,
  velocity dispersion $\sigma$, and higher-order moments $h_3$ and
  $h_4$ of the line-of-sight velocity distribution, respectively. The
  {\it bottom panels\/} show the residuals obtained by subtracting the
  best-fitting model from data.}
\label{fig:fcc167_kin}
\end{figure*}

Fig.~\ref{fig:FCC167_maps} shows the stellar kinematic maps of the
mean velocity $v$, velocity dispersion $\sigma$ and the higher-order
moments $h_3$ and $h_4$ of the line-of-sight velocity distribution for
FCC~167, derived from the $S/N=40$ Voronoi-binned MUSE data. The
presence of a large-scale kinematically cold disc component is
beautifully brought out by the $h_3$ and $h_4$ maps.  The median
errors in the maps are of 5 and 6 km~s$^{-1}$ for $v$ and $\sigma$,
respectively and of $0.025$ for both $h_3$ and $h_4$. To further
reduce these errors for the purpose of constructing dynamical models,
the data were binned to a $S/N=100$ target (see
Fig.~\ref{fig:fcc167_kin}).  This brings down the previous errors on
$v$ and $\sigma$ to 2.5 and 3.5 km~s$^{-1}$ and to 0.012 for $h_3$
and $h_4$. The formal uncertainties on the stellar kinematic
parameters returned by pPXF were rescaled assuming that the $\chi^2$
of the best-fitting model matches the number of degrees of
freedom. The rescaled uncertainties capture the systematic deviations
due to template mismatch and spatial correlations in the Voronoi bins
in the centre and outskirts of the sample galaxies, respectively.

The stellar kinematic maps obtained for FCC~167 with the $S/N=100$
Voronoi-binned MUSE data were used as constraints for Schwarzschild
orbit-based dynamical models in order to derive the internal orbital
structure using the approach described in \citet{vandenBosch2008} and
\citet{Zhu2018a}. The observed surface-brightness distribution of
FCC~167 was fitted with a bi-axial multi-Gaussian-expansion model
\citep{Emsellem1994, Cappellari2002}, and deprojected to a
three-dimensional triaxial model for the stellar luminosity by
adopting a set of viewing angles ($\vartheta, \psi,
\phi$). Multiplication by a constant stellar $M/L$ provides the
intrinsic mass density distribution of the stars.  The three viewing
angles relate directly to the axis ratios $p=b/a$ and $q=c/a$ and the
scale factor $u=a_{\rm obs}/a$, where $a \geq b \geq c$ are the
semi-axes of the constituent triaxial Gaussian density distributions
and $a_{\rm obs}$ is the observed major axis scale-length. Varying $p$
and $q$ allows exploring different intrinsic triaxial shapes. A
spherical dark matter halo was added with a NFW profile
\citep{Navarro1997} and concentration parameter $c$ fixed according to
its relation with the virial mass $M_{200}$ from cosmological
simulations \citep{Dutton2014}. The resulting total mass distribution
provides the gravitational potential. Thus, the dynamical models have
five free parameters: the stellar $M/L$; the axis ratios $p$ and $q$
and scale factor $u$ of the triaxial stellar density distribution and
the dark halo mass $M_{200}$. The gravitational potential is static
and figure rotation is not included in the model.

For each choice of parameters, a large number of stellar orbits was
computed to be able to match the data. The number of orbits sampled in
the space of the three integrals of motion $(E, I_2, I_3)$ was $(65,
15, 15)$, with an additional overall factor $3^3$ resulting from
dithering to create an orbit bundle \citep[see][]{vandenBosch2008},
leading to a total of 394875 orbits per model. This large number of
orbits is needed to do justice to the high quality of the data. The
orbital weights were determined by simultaneously fitting the
orbit-superposition models to the projected and de-projected
luminosity density and the two-dimensional line-of-sight stellar
kinematics, i.e., the maps of $v$, $\sigma$, $h_3$, and $h_4$. As
shown in Fig.~\ref{fig:fcc167_kin}, the model with the best-fitting
global parameters stellar $M/L=5.5$ M$_\odot$/L$_\odot$, $q=0.60$,
$p=0.965$, $u=0.966$ (which corresponds to an inclination angle of
$77.6$ degrees), and $M_{200}=6.4 \times 10^{12}$ M$_\odot$ matches
the stellar kinematic maps well from the inner to the outer regions.

\begin{figure*}[t!]
\centering
\includegraphics[width=\hsize]{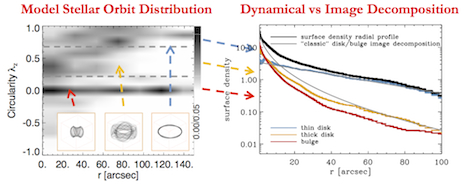}
\caption{Orbital and photometric decomposition of FCC~167. {\it Left
    panel:\/} Orbit distribution in the phase space of circularity
  $\lambda_Z$ versus intrinsic radius $r$, as derived from the
  best-fitting orbit-superposition model shown in
  Fig.\ \ref{fig:fcc167_kin}. Grey-scale colours indicate the orbital
  density in phase space. The horizontal dashed lines separate stars
  in highly-circular cold thin-disc orbits ($\lambda_Z>0.7$) from
  those following dynamically warm thick-disc motions ($0.2 <
  \lambda_Z < 0.7$) and those in the non-rotating bulge and stellar
  halo ($\lambda_Z \sim 0$). {\it Right panel:\/} Stellar
  surface-density profiles of FCC~167. The grey lines trace the
  profiles of a classical exponential disc and S\'ersic bulge as
  derived from a standard bulge-disc decomposition, whereas the blue,
  orange, and red lines follow the projected surface brightness of the
  dynamically cold, warm (thick disc), and hot (bulge) components,
  respectively.}
\label{fig:fcc167_dynmodel}
\end{figure*}

\begin{figure*}[t!]
\includegraphics[width=9.5cm]{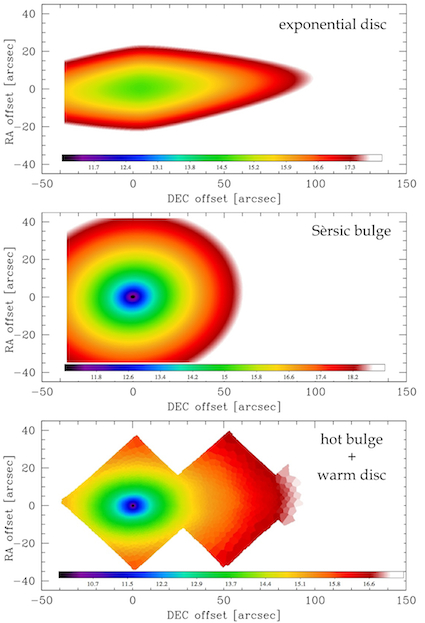}
\includegraphics[width=9.5cm]{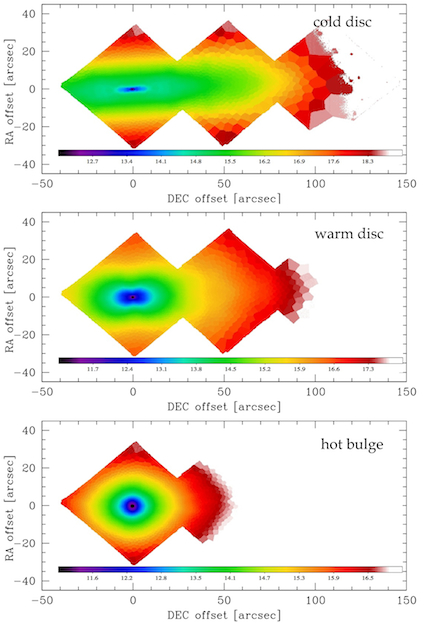}
\caption{Two-dimensional light distribution for the dynamical and photometric components of FCC~167. 
{\it Left panel:\/} Two-dimensional distribution of light from each of the dynamical components, 
    as derived from the best-fitting orbit-superposition model shown in
  Fig.~\ref{fig:fcc167_kin}: dynamically cold, warm (thick disc), and hot (bulge) 
  components are shown in the top, middle and lower panel,
  respectively.  {\it Right panel:\/} Two-dimensional distribution of light from each of the photometric components, 
  S\'ersic bulge (middle panel) and exponential disc (top panel), 
  as derived from the fit of the light distribution of the MUSE reconstructed image shown in
  Fig.~\ref{fig:quality_photometry}. 
  In the lower panel is shown the resulting image from the superposition of the hot (bulge) 
  and warm (thick disc) dynamical components.
  In all panels, the colour bar indicates the surface brightness levels adopting an arbitrary zero point.}
\label{fig:fcc167_2Dmodel}
\end{figure*}

The internal dynamical structure of each model is described by the
distribution of orbital weights in integral-of-motion space. One way to
visualise it is to plot the variation of $\lambda_z$ with radius $r$,
where $\lambda_z$ is the circularity of orbit defined as the angular
momentum around the short axis $z$ normalised by that of the circular
orbit at the same binding energy, and $r$ is the time-averaged radius
of each orbit. The result for the best-fitting model of FCC~167 is
shown in the left panel of Fig.~\ref{fig:fcc167_dynmodel}. Most orbits
have $\lambda_z<0.25$, comprising the bulge in the inner region, and
contributing to the halo in the outer region. The diagram also reveals
two embedded disc components: a warm ($0.25<\lambda_z<0.8$) component
and a cold ($0.8<\lambda_z<1.0$) component. The cold component
resembles a thin disc similar to those seen in spiral galaxies. The
warm orbits constitute a thick disc with weaker rotation.
{The surface brightness profiles of the hot, warm, and cold orbital components
are shown in the right panel of Fig~\ref{fig:fcc167_dynmodel} and the two-dimensional surface 
brightness distributions of these three components are shown in the left panels of  Fig.~\ref{fig:fcc167_2Dmodel}.
The right panels of Fig.~\ref{fig:fcc167_2Dmodel} show the result of a
standard two-dimensional photometric decomposition into a bulge and
disc, performed by using the GALFIT package \citep{Peng2010}
on the MUSE reconstructed image of FCC~167 (see Fig.~\ref{fig:fcc167_kin}).
The surface brightness of the cold component matches the
photometric disc, with comparable scale length ($\sim 30$~arcsec or $\sim3$~kpc).
The photometric bulge appears more boxy than the hot dynamical bulge, but both components have
comparable effective radius ($\sim 18$~arcsec or $\sim2$~kpc).
The light distribution of the combined warm and hot component (see lower-right panel of Fig.~\ref{fig:fcc167_2Dmodel}) 
matches quite well the isophotes of the photometric bulge. This demonstrates that
the dynamical decomposition is needed to disentangle the warm and hot components.}

\subsection{Photometric and kinematic disc signatures in ETGs}
\label{sec:kin}

\begin{figure}[t!]
\centering
\includegraphics[width=\hsize]{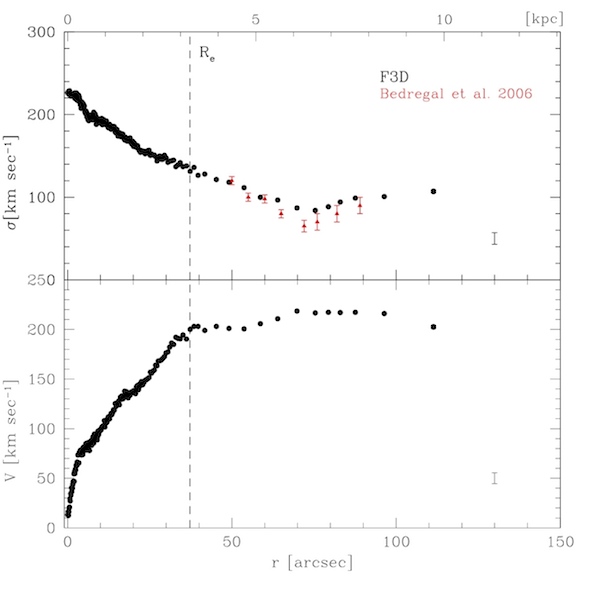}
\caption{Stellar kinematics of FCC~167 extracted from the MUSE data
  (black points) at $\rm P.A.=3.9$ degrees within the 2-arcsec wide
  aperture crossing the nucleus shown in
  Fig.~\ref{fig:quality_photometry}. The mean velocity $v$ ({\it
    bottom panel\/}) and velocity dispersion $\sigma$ ({\it top
    panel\/}) are given. The mean errors of the data are shown in the
  lower-right corner of each panel. The velocity dispersion is
  compared to the long-slit spectroscopic data obtained by
  \citet{Bedregal2006} for the north side of the galaxy (red
  points). The vertical dashed line corresponds to the effective
  radius.}
\label{fig:fcc167_vsigma}
\end{figure}

The radial profiles of the mean velocity and velocity dispersion of
FCC~167 were derived within a 2-arcsec wide aperture at ${\rm P.A.}  =
3.9$ degrees crossing the nucleus (Fig.~\ref{fig:fcc167_vsigma}) from
the kinematic maps obtained after Voronoi binning to a $S/N =
40$. MUSE data map the stellar kinematics out to the unprecedented
galactocentric distance of 115 arcsec (about $11.8$~kpc) with very low
uncertainties ($\leq 10$ km~s$^{-1}$) and definitely confirm the
presence of several localised substructures, which were previously
identified by \citet{Bedregal2006}.

The mean velocity reaches a maximum of $v \sim 200$ km~s$^{-1}$ at
$R\sim50$ arcsec and it remains almost constant out to $R\sim100$
arcsec. The velocity dispersion decreases from a central maximum value
of $\sigma \sim 240$ km~s$^{-1}$ to $\sigma \sim 90$ km~s$^{-1}$ at
$R\sim70$ arcsec. At larger radii, the velocity dispersion increases
again to $\sigma\sim110$ km~s$^{-1}$ for $70 \leq R \leq 115$
arcsec. Such a local minimum is already visible in long-slit data by
\citet{Bedregal2006}, although they did not comment on it. The
$\sigma$-drop seems to be more pronounced in the long-slit data
($\sigma \sim 85$ km~s$^{-1}$ at $R\sim70$ arcsec), but it is
consistent within the errors with the MUSE measurements.

For $70\leq R \leq 115$ arcsec, light is still mapping the bright
regions of the disc in the range of surface brightness $21 \leq \mu_r
\leq 24$ mag~arcsec$^{-2}$ (Fig.~\ref{fig:quality_photometry}). The
$\sigma$-drop corresponds to a plateau in the surface-brightness
radial profile at $R\sim70$ arcsec. At the same radius, the isophotes
show an increasing flattening ($\epsilon \sim0.6$) and a twisting of
10 degrees \citep{Caon1994}.

\begin{figure}[t!]
\centering
\includegraphics[width=5cm, angle=-90]{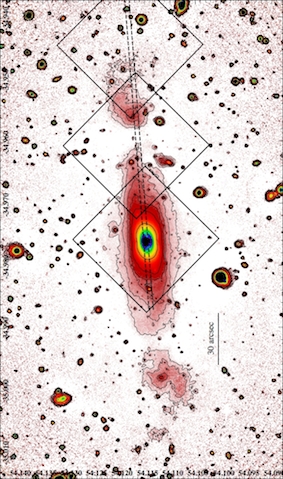}
\caption{High-frequency residual image obtained from the $r$-band VST
  image of FCC~167.  The $1 \times 1$ arcmin$^2$ MUSE pointings are
  overplotted in black.  The black dashed lines mark the 2-arcsec wide
  aperture at $\rm P.A. = 3.9$ degrees, where the stellar kinematics
  shown in Fig.~\ref{fig:fcc167_vsigma} was extracted from the MUSE
  data. The right ascension and declination (J2000.0) are given in
  degrees on the vertical and horizontal axes of the field of view,
  respectively.}
\label{fig:fcc167_fmedian}
\end{figure}

A Gaussian filter of the $r$-band VST image of FCC~167 given in
Fig.~\ref{fig:pointings_fds} was performed by using the
IRAF\footnote{The Image Reduction and Analysis Facility is distributed
  by the National Optical Astronomy Observatory (NOAO), which is
  operated by the Association of Universities for Research in
  Astronomy (AURA), Inc. under cooperative agreement with the National
  Science Foundation.} task FMEDIAN. The ratio between the original
and filtered image gave the high-frequency residual image shown in
Fig.~\ref{fig:fcc167_fmedian}.  In the inner $\sim60$ arcsec, it
reveals a boxy structure which ends with a spiral-like arm and a
bright knot at $R\sim70$ arcsec on both sides along the major axis of
the disc. This is the distance where the $\sigma$-drop in kinematics
and break in the surface-brightness radial profile are observed. Thus,
it seems reasonable to conclude that this feature is more likely a
substructure in the disc, which produces a broadening of the
line-of-sight velocity distribution.

\subsection{Stellar population in the outskirts of ETGs}
\label{sec:pops}

One of the ambitious goals of the F3D project is to study the stellar
age, metallicity, and $\alpha$-element abundance in the outskirts of
ETGs, out to the faint regions of their stellar halos. This will
provide insight into the assembly history of galaxies.

\begin{figure}[t!]
\centering
\includegraphics[width=\hsize]{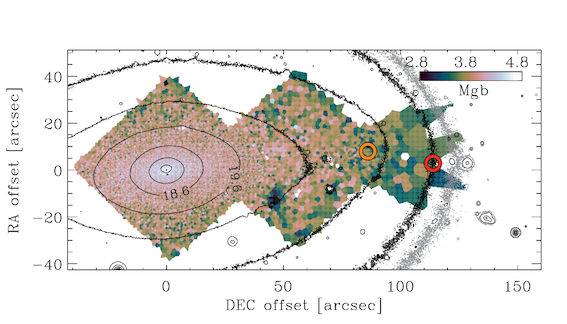}
\caption{{Mg$\,b$} map of FCC~167 with the $r$-band isophotes (black
  lines) and $B-$band isophote (grey line) shown in
  Fig.~\ref{fig:quality_photometry}.  The orange and red circles
  indicate the location of two selected apertures where the
  H$\beta_o$, {Mg$\,b$}, Fe5270, Fe5335, and Fe5406 line-strength
  indices were measured and fitted in order to constrain the stellar
  population properties, as shown in Fig.~\ref{fig:fcc167_MCMCfit}.}
\label{fig:fcc167_Mgb}
\end{figure}

\begin{figure*}[t!]
\centering
\includegraphics[width=9cm]{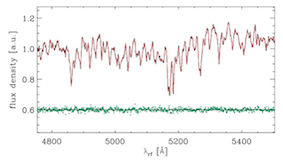}
\includegraphics[width=9cm]{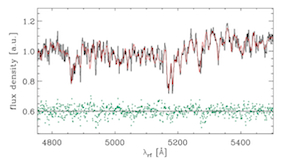}
\includegraphics[width=9cm]{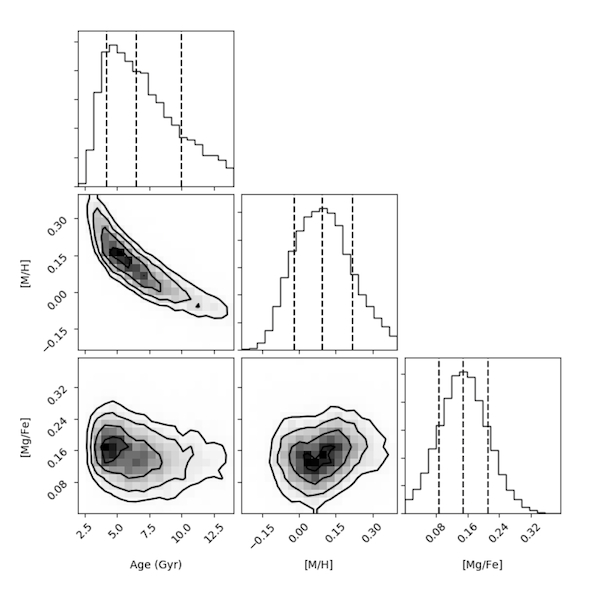}
\includegraphics[width=9cm]{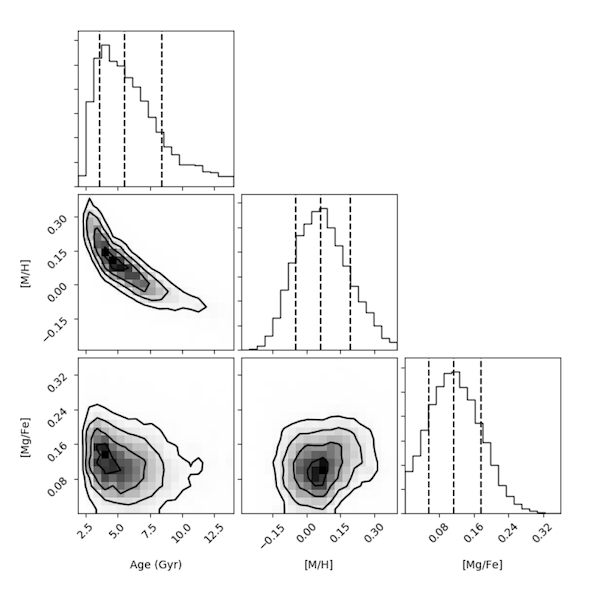}
\caption{Stellar population properties from the two selected apertures
  in the middle ({\it left panels\/}) and halo pointing ({\it right
    panels\/}) of FCC~167 shown in Fig.~\ref{fig:fcc167_Mgb}. {\it Top
    panels:\/} Rest-frame spectra (black lines), best-fitting models
  (red lines), and residuals obtained by subtracting models from
  spectra (green points).  {\it Bottom panels\/:} Posterior
  distributions for the age, metallicity [M/H], and $\alpha$-element
  abundance [Mg/Fe]. The dashed lines overplotted to the histograms
  mark the $16\%$, $50\%$, and $84\%$ percentiles of the marginalised
  distributions.}
\label{fig:fcc167_MCMCfit}
\end{figure*}

Fig.~\ref{fig:fcc167_Mgb} presents the map of the {Mg$\,b$} index,
which shows a clear negative outwards gradient towards lower
metallicity, and reveals the presence of low-metallicity objects along
the line-of-sight of FCC~167 at about 40 arcsec slightly northwest to
the nucleus. The line-strength indices of H$\beta$, {Mg$\,b$},
Fe5270, Fe5335, and Fe5406 were measured in the two selected apertures
in the middle ($\mu_r \sim 21.3$ mag~arcsec$^{-2}$) and halo pointing
($\mu_r \sim 23$ mag~arcsec$^{-2}$) shown in
Fig.~\ref{fig:fcc167_Mgb}. They were fitted with a
single-stellar-population Markov-Chain-Monte-Carlo code, using the
MILES models \citep{Vazdekis2010} to provide the stellar age,
metallicity, and $\alpha$-element abundance ratio. The results are
shown in Fig.~\ref{fig:fcc167_MCMCfit} and suggest that in the
outskirts of FCC~167 the mean stellar population has an
$\alpha$-element abundance $[{\rm Mg/Fe}] \sim 0.8-0.16$ dex and an
average age in the range 2.5--5 Gyr. These initial results are based
on the standard data reduction, and demonstrate that the stellar
population properties can be traced well into its outskirts.

\subsection{Globular clusters and planetary nebulae}
\label{sec:GCs_PNe}

The MUSE observations also give access to GCs and PNe as discrete
tracers of the underlying gravitational potential of the galaxies.
First representative results on the detection of GCs and PNe in
FCC~167 are presented below.

{\it GCs detection in FCC~167 -\/} The IMFIT algorithm
\citep{Erwin2015} was used to model the surface-brightness
distribution of FCC~167 extracted from a broad combined wavelength
range of the MUSE datacubes. The resulting image was subtracted to
highlight underlying structures and point sources.  The locations of
GCs were determined with DAOStarFinder, a Python implementation of the
DAOPHOT algorithm \citep{Stetson1987} and were cross-referenced with
the positions of GC candidates from the HST/ACS GC catalogue
\citep{Jordan2015}.  For FCC~167, the ACS study did not extend into
the outer halo pointing, so additional GC candidates were identified
directly from the MUSE data.  The redshifts of these additional GC
candidates were checked to confirm that these are really GCs in the
FCC~167 halo rather than stars or background galaxies.

\begin{figure*}[t!]
\centering
\includegraphics[width=\hsize]{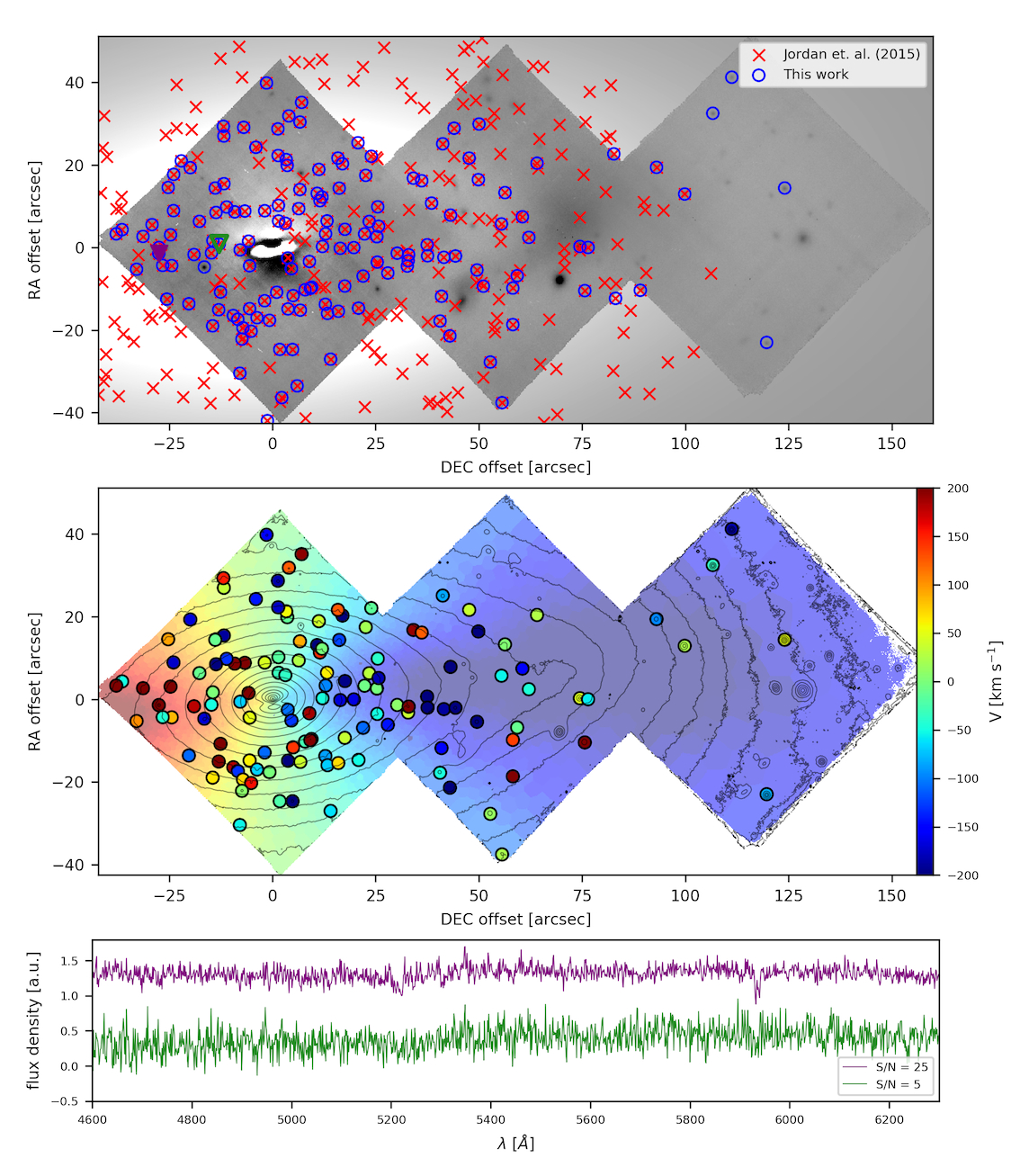}
\caption{Illustration of GC extraction from MUSE data of FCC~167.
  {\it Top panel:\/} Residual image obtained by subtracting the IMFIT
  surface-brightness model of FCC~167 from the galaxy
  surface-brightness distribution extracted from MUSE data. The GC
  candidates from MUSE data (blue circles) and GCs in the HST/ACS GC
  catalogue \citep[][red crosses]{Jordan2015} are shown. {\it Middle
    panel:\/} Map of the mean stellar velocity of FCC~167 shown in
  Fig.~\ref{fig:fcc167_kin} and radial velocities of the GCs. {\it
    Bottom panel:\/} Spectra of two MUSE GCs with $S/N\approx25$
  (purple line) and $5$ (green line) whose location is indicated in
  the top panel by the filled purple and open green triangles,  respectively.}
\label{fig:fcc167_GCs}
\end{figure*}

The seeing of the observations for FCC~167 was about 0.9 arcsec (FWHM)
corresponding to $\sigma=0.38$ arcsec (about 2 pixels). Therefore, the
spectra were extracted within a circular aperture with a radius of two
pixels. The contamination from the galaxy light was removed by
subtracting the galaxy spectrum in an annulus around the GC. The
radial velocities of the GCs were determined via full spectral fitting
with pPXF. Fig.~\ref{fig:fcc167_GCs} shows the map of the mean stellar
velocity of FCC~167 after Voronoi binning to a $S/N = 100$ with
extracted GCs colour-coded by their radial velocities. While the
stellar population properties can be determined only for GCs with high
$S/N$, the low $S/N$ of the majority of detected GCs still allows for
radial velocity estimates.

\begin{figure}[t!]
\centering
\includegraphics[width=\hsize]{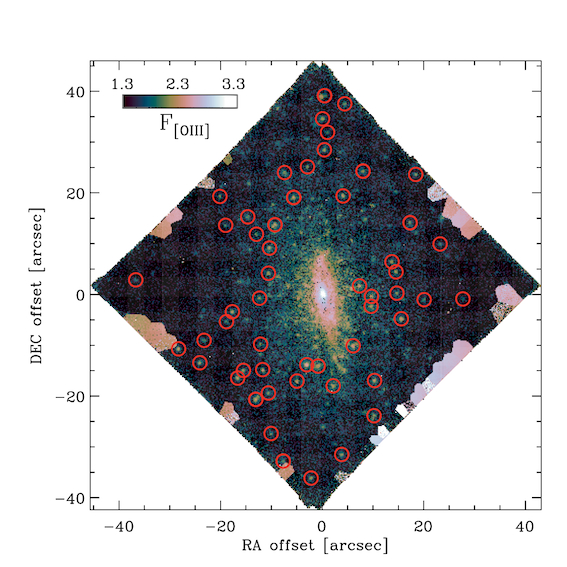}
\caption{Map of the flux of the [\ion{O}{iii}]$\lambda5007$ emission
  from the central pointing of FCC 167. The extended disc is clearly
  visible. The detected planetary nebulae are indicated by red
  circles.  [\ion{O}{iii}]$\lambda5007$ flux values are shown on a
  logarithmic scale and in $10^{-20}$ erg~cm$^{-2}$~s$^{-1}$.}
\label{fig:fcc167_PNe}
\end{figure}

{\it PNe in FCC~167 -\/} To illustrate the potential of the MUSE
observations to detect and study the PNe in the central regions of
galaxies out to the distance of the Fornax cluster,
Fig.~\ref{fig:fcc167_PNe} shows a map of the flux of the
[\ion{O}{iii}]$\lambda5007$ emission from the central pointing of
FCC~167 obtained when seeing conditions were optimal. In addition to a
central disc of diffuse ionised gas \citep{Viaene2018}, the presence
of a large number of point sources is immediately evident. The most
prominent ones are highlighted in Fig.~\ref{fig:fcc167_PNe} and are
confirmed as PNe.  Fig.~\ref{fig:fcc167_PNe} also shows how it becomes
progressively more difficult to establish the presence of PNe towards
the central regions. This is due to an increasing background of
[\ion{O}{iii}]$\lambda5007$ emission that consists, in fact, mostly of
correlated false detections. As the amplitude of false detections
scales with the Poisson noise level in the stellar continuum of the
MUSE spectra, this increase in the spurious emission towards the
centre of the galaxy makes it more difficult to identify fainter PNe.

A preliminary analysis of the PNe population in FCC~167 indicates that
PNe can be detected down to the absolute [\ion{O}{iii}]$\lambda5007$
magnitude $M_{5007}= -2.7$, or nearly two magnitudes below the bright
cutoff of the PNe luminosity function \citep{Ciardullo1989}, while
being complete at $M_{5007} = -3.7$. A detailed investigation of the
PNe population of this and other F3D galaxies is under
way. Furthermore, the unique depth of the MUSE stellar-population
measurements will allow to check the connection between the PNe
properties and those of their parent stellar populations
\citep[e.g.][]{Buzzoni2006} both {\it a)\/} in a spatially consistent
fashion and {\it b)\/} far out into the outskirts of F3D galaxies. In
these outer regions narrow-band or slitless spectroscopic measurements
from the literature \citep[see, e.g.,][for specific Fornax
  measurements]{Spiniello2018} will complement well the measurements
from the F3D data.

\subsection{Late-type galaxies}
\label{sec:LTGs}

\begin{figure*}
\begin{center}
\includegraphics[width=\hsize]{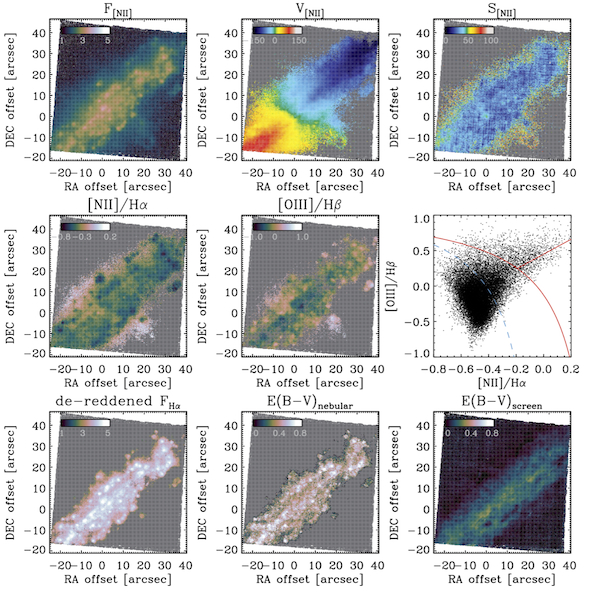}
\end{center}
\caption{Emission-line analysis for the central regions of FCC~312.
  {\it Top panels:\/} Maps of the total flux, mean velocity and
  velocity dispersion along the line-of-sight from the
  [\ion{N}{ii}]$\lambda6583$ nebular emission.  {\it Middle panels:\/}
  Maps of the [\ion{N}{ii}]$\lambda6583$/H$\alpha$ and
  [\ion{O}{iii}]$\lambda5007$/H$\beta$ line ratios and corresponding
  BPT diagram.  {\it Bottom panels:} Maps of the de-reddened H$\alpha$
  flux, extinction on the nebular lines, and extinction derived from
  the stellar continuum.  Velocities and velocity dispersions are in
  $\rm km\,s^{-1}$, reddening values are in magnitude, line ratios and
  fluxes are shown in a logarithmic scales, with the latter in units
  of $10^{-20}$ erg~cm$^{-2}$~s$^{-1}$.  Grey areas in the maps refer
  to regions where the $S/N$ ratio of the relevant lines was not
  sufficiently elevated to firmly exclude a false positive
  detection. These levels were established in regions devoid of
  emission well above and below the equatorial plane of the galaxy.}
\label{fig:fcc312_emission}
\end{figure*}

Fig.~\ref{fig:fcc312_emission} shows initial results of the emission
line analysis performed with pPXF and GandALF on the late-type galaxy
FCC~312. This is an Scd galaxy on the east side of the cluster, at
about $2$ degrees from NGC~1399. It is the most extended disc galaxy
in the F3D sample, with a major axis diameter of 4.7 arcmin, with a
prominent flare on both sides of the outer disc regions. The three
MUSE pointings map the light from the centre out to the flaring
regions of the disc, which are more pronounced on the southeast side
(Fig.~\ref{fig:pointings_dss}). Here, initial results are reported for
the central pointing. The ten LTGs in the sample, which span the full
range in the Hubble sequence of barred and unbarred spirals, from Sa
to Sd and Sm types, will be presented and discussed in a forthcoming
paper.

The top row of Fig.~\ref{fig:fcc312_emission} shows the maps of the
total flux from the [\ion{N}{ii}]$\lambda6583$ emission and of the
corresponding velocity and velocity dispersion. The flux map is
characterised by the conspicuous presence of regions with gas ionised
by hot stars, and, interestingly, a loop-like feature in the southeast
side suggestive of expanding gaseous bubbles produced by stellar
feedback.  While the velocity dispersion map does not show an excess
in the loop, which could originate from shocks with the cold phase of
the interstellar medium produced by the outward motion, the velocity
field shows an offset in the region as compared to the rest of the
galaxy. The [\ion{N}{ii}]$\lambda6583$/H$\alpha$ and
[\ion{O}{iii}]$\lambda5007$/H$\beta$ line ratio maps displayed in
Fig.~\ref{fig:fcc312_emission} also present conspicuous extra-planar
ionised regions in the area of the loop-like feature, as well as
significant variation in the [\ion{O}{iii}]$\lambda5007$/H$\beta$ line
ratio amongst the different \ion{H}{ii} regions. It is possible that
this feature is another example of an extra-planar echo of activity in
the galactic nucleus \citep[see][]{Keel2012}. The spaxel-by-spaxel BPT
diagram \citep{Baldwin1981} of Fig.~\ref{fig:fcc312_emission} shows
that several line-excitation mechanisms are present, including that
from AGN activity.

The H$\alpha$ de-reddened flux map is shown in the bottom row of
Fig.~\ref{fig:fcc312_emission}, together with maps of the dust
extinction, as measured in the nebular lines and from the stellar
continuum with the usual uniform dust screen approximation across the
galaxy. The latter modifies the entire spectrum, and is thus in
addition to the extinction affecting nebular lines only. The nebular
extinction map is derived using the standard Balmer decrement
approach, and accounting for the uniform dust screen
contribution. This differentiated approach to measure the extinction
allows deriving more accurately the effects of dust mixed with gas in
star forming regions. Comparing the extinction maps in
Fig.~\ref{fig:fcc312_emission} one sees that the continuum-derived
dust component in the galaxy is more uniformly distributed and
concentrated in the disc plane, as expected.\looseness=-2

\begin{figure}
\begin{center}
\includegraphics[width=\hsize]{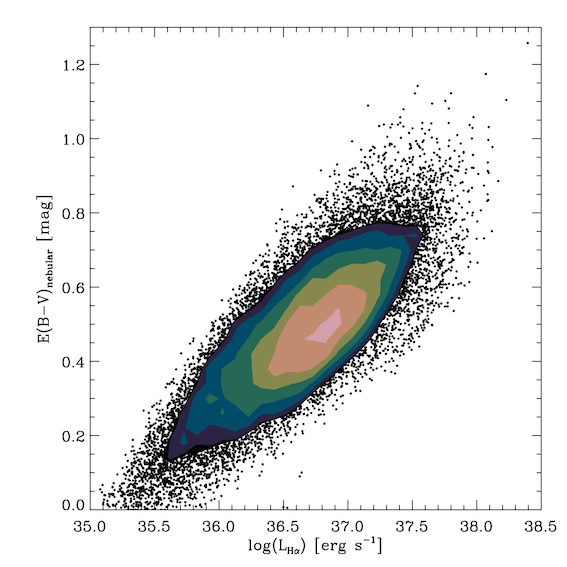}
\end{center}
\caption{Correlation between the dust extinction affecting nebular
  lines and H$\alpha$ luminosity in the same region, corrected for
  total extinction.  Regions with elevated star formation rate are
  also heavily dust obscured.}
\label{fig:fcc312_dustysf}
\end{figure}

Interestingly, the regions with elevated H$\alpha$ flux (and thus star
formation rate) are more heavily dust-obscured. To illustrate this,
Fig.~\ref{fig:fcc312_dustysf} shows the correlation found between the
dust extinction affecting nebular lines and H$\alpha$ luminosity in
the same region. The origin of this correlation can be understood if
regions of elevated star formation are also the regions of more
substantial gas content, which is to be expected from the
Kennicutt-Schmidt star-formation relation \citep{Schmidt1959,
  Kennicutt1998}. More ionising photons coming from hot stars combined
with more hydrogen atoms would lead to more luminous H$\alpha$
emission. Furthermore, more gas can in principle be translated into
more dust content as well, but this correlation can depend on the
metal content of the gas, since metal-poor regions are inefficient in
producing dust grains. This can partly explain the scatter in the
correlation. Another factor that can add scatter is the geometry of
the dust distribution in the individual star-forming systems, which
will determine the fraction of escaping photons. A study of this
correlation for all F3D LTGs will be presented in a future paper.

\begin{figure}
\begin{center}
\includegraphics[width=\hsize]{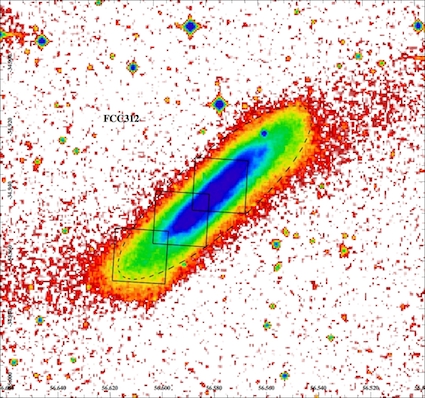}
\end{center}
\caption{Digitized Sky Survey image of FCC~312. The $1 \times 1$
  arcmin$^2$ MUSE pointings are shown in black. The black dashed
  ellipse corresponds to the isophote at $\mu_B=25$ mag~arcsec$^{-2}$,
  while the right ascension and declination (J2000.0) are given in
  degrees on the horizontal and vertical axes of the field of view,
  respectively.}
\label{fig:fcc312_sdss}
\end{figure}

\section{Concluding remarks}
\label{sec:remarks}

This paper has demonstrated that MUSE enables a comprehensive study of
the internal orbital structure and stellar populations of a sample of
galaxies inside the virial radius of the Fornax cluster. The design of
the survey, the selection of the sample, the observations and the main
steps of the data analysis were described in Sects.~\ref{sec:plan},
\ref{sec:obs}, and \ref{sec:analysis}, respectively. The assessment of
the data quality in Sect.~\ref{sec:quality} and initial results
included in Sect.~\ref{sec:results}, obtained for the early-type
galaxy FCC~167 and the late-type galaxy FCC~312, confirm that the
goals of the project as outlined in the Introduction can be achieved.

Specifically, the results for FCC~167 demonstrate that the MUSE data
allow reliable extraction of the two-dimensional stellar kinematic
maps of the mean velocity $v$, velocity dispersion $\sigma$, and
higher-order moments $h_3$ and $h_4$ of the line-of-sight velocity
distribution, and of the key line-strength indices to a surface
brightness of $\mu_B \sim 25$ mag~arcsec$^{-2}$, i.e., to the
outskirts of the galaxies in the Fornax cluster. Orbit-based dynamical
models reveal that FCC~167 contains not one, but two embedded discs,
one thin and the other thick. The former is also seen photometrically,
but the latter is not. Stellar population models applied to the
line-strength measurements provide ages, metallicities and
$\alpha$-element abundances, and possibly the IMF, well into the halo
region.

The emission-line maps of total flux, mean velocity and velocity
dispersion for FCC~312 provide detailed distribution of the regions
ionised by hot stars, and reveal a loop-like feature suggestive of an
expanding gas bubble. The nebular lines provide an estimate of the
dust extinction and delineate regions of star formation. The sample
contains ten such objects.

The MUSE data also allows detection and characterisation of globular
clusters. These objects provide an additional tracer of the formation
history of the galaxies. Individual planetary nebulae can be detected
as well through their [\ion{O}{III}]$\lambda5007$ emission, including
in the bright inner regions. The central pointing also provides a
unique census of the nuclear star clusters across different galaxies
in the Fornax cluster.

Work is underway along these and other lines of study, and will be
reported in future papers.


\begin{acknowledgements}
Based on observations collected at the European Organisation for
Astronomical Research in the Southern Hemisphere under ESO programme
296.B-5054(A).
It is a pleasure to thank Roland Bacon and the MUSE team for building such a marvellous instrument, 
to thank the staff at ESO who expertly carried out
and supported the service observations, and to thank Francesco La
Barbera, Lorenzo Morelli, Borislav Nedelchev, Francesca Pinna, Adriano
Poci, and S\'ebastien Viaene for their comments.
The F3D team is grateful to the P.I.s of the Fornax Deep Survey (FDS)
with VST (R.F.\ Peletier and E.\ Iodice) who kindly provided the
thumbnail of the FDS mosaic in the $r$ band centred on FCC~167.
E.M.C. and A.P. acknowledge financial support from Padua University
through grants DOR1715817/17, DOR1885254/18 and BIRD164402/16.
J.F.-B. acknowledges support from grant AYA2016-77237-C3-1-P from the
Spanish Ministry of Economy and Competitiveness
(MINECO). G.v.d.V. acknowledges funding from the European Research
Council (ERC) under the European Union's Horizon 2020 Research and
Innovation Programme under grant agreement No. 724857 (Consolidator
Grant ArcheoDyn).
R.McD. is the recipient of an Australian Research Council Future Fellowship (project number FT150100333).
This research made use of the Digitized Sky Surveys produced at the
Space Telescope Science Institute, USA
(http://archive.stsci.edu/dss/), HyperLeda Database maintained by the
Observatoire de Lyon, France and Special Astrophysical Observatory,
Russia (http://leda.univ-lyon1.fr/), and NASA/IPAC Extragalactic
Database (NED) which is operated by the Jet Propulsion Laboratory,
California Institute of Technology, USA
(http://ned.ipac.caltech.edu/).
{\bf The F3D team is indebted to the anonymous referee for a constructive report that was delivered very rapidly, and led
to an improvement of the paper.}

\end{acknowledgements}

\bibliography{f3d}


\begin{appendix}

\section{Digitized Sky Survey images and MUSE pointings of F3D galaxies}
\label{sec:pointings}

\begin{figure*}[t!]
\centering
\includegraphics[width=17cm]{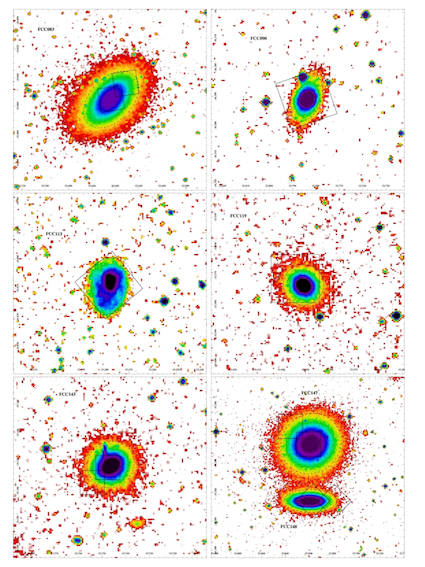}
\caption{Digitized Sky Survey images of the F3D galaxies. For each
  galaxy, the $1 \times 1$ arcmin$^2$ MUSE pointings are shown in
  black. The black dashed ellipse corresponds to the isophote at
  $\mu_B=25$ mag~arcsec$^{-2}$. The right ascension and declination
  (J2000.0) are given in degrees on the horizontal and vertical axes
  of the field of view, respectively.}
\label{fig:pointings_dss}
\end{figure*}

\addtocounter{figure}{-1}
\begin{figure*}[t!]
\centering
\includegraphics[width=17cm]{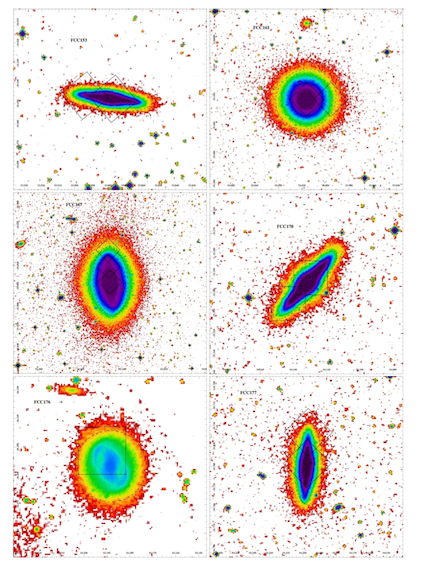}
\caption{continued.}
\end{figure*}

\addtocounter{figure}{-1}
\begin{figure*}[t!]
\centering
\includegraphics[width=17cm]{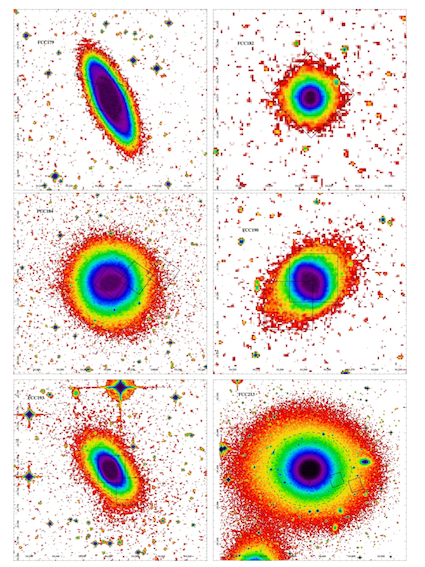}
\caption{continued.}
\end{figure*}

\addtocounter{figure}{-1}
\begin{figure*}[t!]
\centering
\includegraphics[width=17cm]{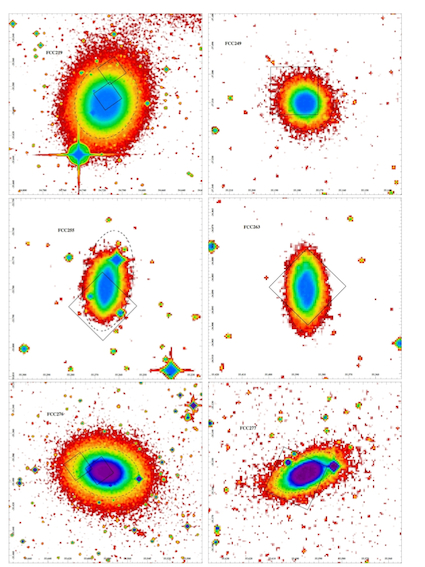}
\caption{continued.}
\end{figure*}

\addtocounter{figure}{-1}
\begin{figure*}[t!]
\centering
\includegraphics[width=17cm]{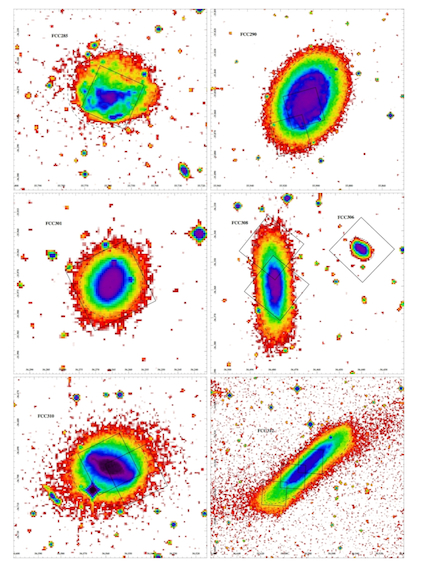}
\caption{continued.}
\end{figure*}

\end{appendix}

\end{document}